\documentclass[reprint,preprintnumbers,aps,prd,amsmath,amssymb,nobibnotes,nofootinbib,onecolumn]{revtex4-2}
\usepackage{xcolor}
\usepackage{subcaption}
\usepackage{amsmath}
\usepackage{amssymb}
\usepackage{graphicx}
\usepackage{hyperref}

\newcommand{\x}{\mathbf{x}}
\newcommand{\y}{\mathbf{y}}

\newcommand{\p}{\mathbf{p}}
\newcommand{\q}{\mathbf{q}}

\begin{document}

\preprint{FERMILAB-PUB-26-0177-T}
\title{Generative models on phase space}

\author{Zachary Bogorad}
\affiliation{Theory Division, Fermi National Accelerator Laboratory, Batavia, IL, USA}
\author{Ibrahim Elsharkawy}
\affiliation{Department of Physics, University of Toronto and Vector Institute, Toronto, ON, Canada}
\affiliation{NERSC, Lawrence Berkeley National Laboratory, Berkeley, California, USA}
\author{Yonatan Kahn}
\affiliation{Department of Physics, University of Toronto and Vector Institute, Toronto, ON, Canada}
\author{Andrew J.~Larkoski}
\affiliation{American Physical Society, Hauppauge, New York, USA}
\author{Noam Levi}
\affiliation{Tel Aviv University, Tel Aviv, Israel}

\begin{abstract}
\noindent Deep generative models such as diffusion and flow matching are powerful machine learning tools capable of learning and sampling from high-dimensional distributions. They are particularly useful when the training data appears to be concentrated on a submanifold of the data embedding space. For high-energy physics data, consisting of collections of relativistic energy-momentum 4-vectors, this submanifold can enforce extremely strong physically-motivated priors, such as energy and momentum conservation. If these constraints are learned only approximately, rather than exactly, this can inhibit the interpretability and reliability of such generative models. To remedy this deficiency, we introduce generative models which are, by construction, confined at every step of their sampling trajectory to the manifold of massless $N$-particle Lorentz-invariant phase space in the center-of-momentum frame. In the case of diffusion models, the ``pure noise'' forward process endpoint corresponds to the uniform distribution on phase space, which provides a clear starting point from which to identify how correlations among the particles emerge during the reverse (de-noising) process. We demonstrate that our models are able to learn both few-particle and many-particle distributions with various singularity structures, paving the way for future interpretability studies using generative models trained on simulated jet data.
\end{abstract}

\maketitle
\tableofcontents

\section{Introduction}
\label{sec:intro}

Deep generative models have demonstrated impressive success in the realms of natural data, such as images, videos, and language, and are being rapidly adopted in particle physics to replace or complement simulations~\cite{hepmllivingreview}. Part of the success of these models may be explained by the manifold hypothesis~\cite{fefferman2013testingmanifoldhypothesis,bengio2014representationlearningreviewnew,Brahma2016WhyDL}, which postulates that natural data lives on a much lower-dimensional submanifold $\mathcal{M}$ of the ambient space $\mathbb{R}^m$. When applied to generic data, these concepts are excruciatingly difficult to make precise. While it certainly seems intuitive that the manifold of human faces is much lower-dimensional than the embedding space $\mathbb{R}^{196,608}$ of $256 \times 256 \times 3$ pixelated RGB images, it is much harder to specify what that dimension actually is, let alone whether a sample from a trained generative model lies on the data manifold or slightly off it. By contrast, physics data typically live on well-defined manifolds, where confinement to the manifold is a consequence of symmetries (e.g.\ rotations, translation in space and time, scaling transformations), or equivalently conservation laws (angular momentum, energy-momentum, traceless stress-energy tensors, respectively) which have been validated by decades or centuries of experiments. In the realm of theoretical physics, we can make these symmetries exact to arbitrary precision, and generate simulated data which is \emph{exactly} confined to a submanifold embedded in a higher-dimensional ambient space. Despite rapid advances in artificial intelligence (AI) methods, relatively little effort has been devoted to ensuring that the generative process itself (as opposed to the model architecture) respects these symmetries or constraints.

Focusing on the case of high-energy physics data, a collection of $N$ relativistic particles produced from the collision of two initial-state particles of fixed total 4-momentum $p^\mu_{\rm beam}$ is a point $P \equiv \{p_I^\mu \}$ on the manifold $\Pi_N$ of $N$-particle Lorentz-invariant phase space, which can be viewed as a collection of $N$ 4-vectors $p_I^\mu$, where capital Latin letters will label particles. The assumed relativistic invariance of the interactions that produce these particles determines the Lorentz-invariant measure to be 
\begin{equation}
    d\Pi_N = \left (\prod_{I = 1}^N d^4p_I\,\delta(p_I^2 - m_I^2)\,\Theta(p_I^0)\right)\delta^{(4)}\left(p^\mu_{\rm beam} - \sum_{I=1}^N p^\mu_I\right),
\end{equation}
where $m_I$ is the mass of particle $I$. A common situation is to take all final-state particles to be massless, $m_I = 0$, and to take the initial particles to be in the center-of-momentum (CM) frame with unit energy, $p^\mu_{\rm beam} = p^\mu_{{\rm CM}} = (1,0,0,0)$. Integrating out the mass-shell delta functions and setting $p^0_I = |\p_I|$ gives 
\begin{equation}
    d\Pi_N = \left (\prod_{I = 1}^N \frac{d^3 \p_I}{2 |\p_I|}\right)\delta^{(3)}\left(\sum_{I=1}^N \p_I\right)\delta\left(1 - \sum_{I=1}^N |\p_I| \right).
    \label{eq:LIPS}
\end{equation}
We can thus treat $P \equiv \{\p_I \} \in \mathbb{R}^{3N}$, with the $N$-particle phase space manifold $\Pi_N$ a $(3N-4)$-dimensional submanifold embedded in $\mathbb{R}^{3N}$. The simplest distributions encountered in high-energy physics have $N =2$, while modern accelerator experiments routinely deal with $N \sim \mathcal{O}(200)$. A deep generative model trained on this type of high-energy physics data will try to learn a distribution $p_0(P) \equiv p_0(\p_1, \p_2, \dots, \p_N)$ on this manifold, which is proportional to the squared matrix element for the collision process as computed in quantum field theory (QFT).\footnote{Since the symbol $p$ appears in several contexts in this paper, we will endeavor to keep a consistent notation: isolated 4-vectors not appearing in a dot product will always be written $p^\mu$, and 3-momentum magnitudes will always be written $|\p|$, so Latin un-bolded $p$ or $p_0$ will always mean a probability distribution except when we refer to phase space as $p$-space.} The delta functions enforce energy and momentum conservation. 

For concreteness, we focus in this paper on two types of generative models: diffusion and flow matching. Diffusion models~\cite{SohlDickstein:2015diffusion,Song:2019scorematching,Ho:2020ddpm,Song:2021sde} train a neural network (NN) to predict a training data point $\x \sim p_0(\x)$ from a noisy version $\x(\epsilon)$ across various noise scales $\epsilon$ (the \emph{forward process}). In the \emph{reverse process}, one can start with a sample from a pure noise distribution, for example a mean-zero normal distribution with unit covariance $\y \sim \mathcal{N}(0,I)$, and progressively de-noise it to obtain a new sample $\hat{\x} \sim p_0(\x)$. Similarly, flow matching~\cite{Lipman:2023flowmatching,lipman2024flowmatchingguidecode,tong2024improving} learns a time-dependent velocity field to transform a simple prior distribution into the data distribution by integrating an ordinary differential equation (ODE).\footnote{See Ref.~\cite{lai2025principlesdiffusionmodels} for a recent review on deep generative modeling, including diffusion models and the relation between diffusion and flow matching.} The noising process or ODE flow toward the prior gradually inflates the data manifold into the ambient space, allowing the denoising NN or reverse ODE flow to gradually confine the data back to $\mathcal{M}$ during sampling. However, current diffusion or flow matching models trained on high-energy physics data~\cite{Leigh:2023toe,Mikuni:2023dvk,Butter:2023fov,Leigh:2023zle,Buhmann:2023zgc,Butter:2023ira,Sengupta:2023vtm,Quetant:2024ftg,Jiang:2025pil,Mikuni:2025tar,Dreyer:2025zhp,Faroughy:2025mlw,OmniLearned} do not yet generate samples which are confined exactly to the phase space manifold, as would be required by energy-momentum conservation. That said, there are several approaches implementing exact Lorentz covariance in the architecture, for example Refs.~\cite{Gong:2022lye,Bogatskiy:2022czk,Hao:2022zns,Bogatskiy:2023nnw,Spinner:2024hjm,Spinner:2025prg}, as well as techniques for ``softly'' imposing symmetries~\cite{Hebbar:2025adf}. We emphasize, though, that energy-momentum conservation constraints are logically distinct from Lorentz transformation properties; at a group theory level, Lorentz invariance does not imply invariance under the larger Poincar\'{e} group.

In this paper, we introduce a general framework, $q$-space generative modeling, for constructing models which can naturally accommodate the symmetries and structures of $\Pi_N$ and the distributions typically encountered in high-energy physics.\footnote{See also Refs.~\cite{xu2022poisson,xu2023pfgm++} for other examples of diffusion models trained on physical systems with symmetry constraints.} Our construction differs from standard deep generative modeling in the following ways:
\begin{itemize}
\item The data stays exactly confined to the phase space manifold at every step during the probability flow (forward and reverse stochastic trajectories for diffusion models, ODE flow for flow matching models). This ensures that samples from the model at any point on the trajectory satisfy exact energy-momentum conservation.
\item The ``pure noise'' endpoint of the forward diffusion process is not taken to be Gaussian noise in each of the components of $\p_I$, but rather the uniform distribution with respect to the Lorentz-invariant measure, Eq.~(\ref{eq:LIPS}). This ensures that any correlations among the $\p_I$ that emerge during the reverse process can be meaningfully distinguished from a flat distribution which imposes no structure or correlations on phase space. While in principle this can also be implemented in flow matching by taking the prior to be the uniform distribution on phase space, we have not yet successfully trained such models; nonetheless we see no fundamental impediment to doing so.
\item The probability flow is manifestly permutation-equivariant; when combined with a permutation-invariant NN architecture, this permits the generative model to efficiently learn permutation-invariant distributions for point clouds of unordered sets of particles, as would be relevant for jets of quantum chromodynamics (QCD).
\end{itemize}
In an amusing historical resonance, these features are most easily implemented not by adapting the techniques of modern diffusion models on Riemannian manifolds~\cite{2022arXiv220202763D,kawasakiborruat2026diffusionprocessesimplicitmanifolds}, but by leveraging the 40-year-old \texttt{RAMBO}~\cite{Kleiss:1985gy} algorithm for permutation-invariant sampling of $N$-particle massless phase space, which trades the constraints of phase space for a non-uniform distribution in an auxiliary space, $q$-space.\footnote{See also Refs.~\cite{Nachman:2023clf,Heimel:2024wph} for other uses of \texttt{RAMBO} and related algorithms in generative models for high-energy physics.}\footnote{Independent of the application to generative models, the geometry of $N$-particle phase space seems to be a rich playground for further study in the context of machine learning~\cite{Komiske:2020qhg}.} We perform the probability flow in $q$-space in a manner reminiscent of the setup for mirrored Langevin dynamics~\cite{hsieh2020mirroredlangevindynamics}. Our method applies equally well to phase space for any particle number $N$, but we expect it to find the most utility for the high-dimensional, large-$N$ distributions where analytic parametrizations may not be available.

This paper is organized as follows. In Sec.~\ref{sec:Diffusion}, we review the standard deep generative techniques and present our modified $q$-space construction. In Sec.~\ref{sec:LowDim}, we demonstrate that diffusion on phase space is able to learn and sample from two 3-body matrix elements, one smooth and one nearly-singular. For the latter, we use the amplitude for the process $e^+ e^- \to q \bar{q} g$, a matrix element with soft and collinear singularities that serves as a representative example for the kinds of matrix elements one encounters in describing QCD jets. Because 3-particle phase space may be easily visualized by projecting the 5-dimensional data manifold $\Pi_3$ to the Dalitz plane, we can demonstrate both qualitatively (visually) and quantitatively (statistically) that the correct distribution is being learned. In the case of the $q \bar{q} g$ amplitude, despite the fact that the diffusion model output is mismatched from the training data for very low-energy particles, the distribution of the infrared and collinear (IRC)-safe observable $\tau = {\rm min}\{p_I \cdot p_J\}$ away from the unphysical $\tau = 0$ region is learned almost perfectly. In Sec.~\ref{sec:Antenna}, we use our $q$-space construction to learn a matrix element on 10-particle phase space that exhibits the general structure of soft and collinear divergences of multi-particle production in QCD.  Again, we find that the diffusion model correctly learns the $\tau$ distribution away from $\tau = 0$, and the flow matching models are able to learn the entire distribution covering 9 orders of magnitude in $\tau$. We conclude in Sec.~\ref{sec:Conclusions} with a broader perspective on the utility of our construction for improving our understanding of AI, as well as high-energy physics. Details of our numerical experiments and a derivation of the $\tau$ distribution on uniform phase space are given in Appendices~\ref{app:DiffusionModelDetails} and~\ref{app:tauder}, respectively.

\section{Deep generative models in $q$-space}
\label{sec:Diffusion}
We begin with a very brief review of standard diffusion and flow matching models, and then explain how to modify them to satisfy our desired structures.

\subsection{Review of diffusion models}

A diffusion model attempts to learn a distribution $p_0(\mathbf{x})$ from empirical samples $\mathbf{x} \sim p_0$ with $\mathbf{x} \in \mathbb{R}^m$. It does so by exploiting the fact that the Langevin dynamics which diffuse $p_0(\mathbf{x})$ into a pure-noise distribution $p_T(\mathbf{x})$ after $T \gg 1$ timesteps may be reversed with knowledge of the time-dependent \emph{score} function $\mathbf{s}^{(t)}(\mathbf{x}) \equiv \nabla \log p_{t}(\mathbf{x})$ for $0 \leq t \leq T$. Specifically, the stochastic \emph{forward process}
\begin{equation}
   \mathbf{x}_{t+1} = (1 - \gamma_t) \mathbf{x}_t + \sqrt{2\gamma_t} \mathbf{Z}_t, \qquad t = 0, 1, \dots, T
    \label{eq:VanillaForward}
\end{equation}
where $\mathbf{x}_0 \sim p_0$, $\mathbf{Z}_t \sim \mathcal{N}(0, I)$ is isotropic Gaussian noise, and $\gamma_t \in (0,1)$ is a fixed time-dependent noise schedule.\footnote{An alternate parameterization seen in the literature is $\mathbf{x}_{t+1} = \sqrt{1-\beta_t^2}\mathbf{x}_t + \beta_t \mathbf{Z}_t$. Taking $\beta_t = \sqrt{2\gamma_t}$ and expanding for small $\gamma_t \ll 1$, this is equivalent to Eq.~(\ref{eq:VanillaForward}).} For finite time $t$, let the distribution of $\mathbf{x}_t$ be $p_t(\mathbf{x})$. As $\gamma_t \to 0$ and $T \to \infty$, the forward process converges to the pure noise distribution $p_{T\to \infty} \approx \mathcal{N}(0, I)$. In words, the forward process destroys the signal distribution by adding Gaussian noise to each component. Likewise, starting from pure noise, we can recover $p_0$ as the equilibrium distribution as long as we use the score function to push back against the Gaussian noise. Defining the time-reversed quantities $\bar{\mathbf{x}}_{t} = \mathbf{x}_{T-t}$ and $\bar{\gamma_t} = \gamma_{T-t-1}$, the \emph{reverse process}
\begin{equation}
    \bar{\mathbf{x}}_{t+1} = (1+\bar{\gamma}_t)\bar{\mathbf{x}}_t + 2 \bar{\gamma}_t \nabla \log p_{T-t}(\bar{\mathbf{x}}_t) + \sqrt{2\bar{\gamma}_t}\mathbf{Z}_t
    \label{eq:VanillaReverse}
\end{equation}
starting from the initial distribution $\bar{\mathbf{x}}_0 \sim \mathcal{N}(0, I) \approx p_{T \to \infty}$, converges back to $\bar{\mathbf{x}}_{T \to \infty} \sim p_0$. 

In a typical application, $p_{t}(\mathbf{x})$ is not analytically computable and is thus modeled by a NN. A diffusion model uses the forward process to train a network $\mathbf{s}_\theta$ to predict the score, which one can think of as learning to ``denoise'' the data. It can be shown that the $\mathbf{s}^{(t)}_\theta(\mathbf{x})$ which, averaged over the data distribution, best approximates the true score $\nabla \log p_t(\mathbf{x})$, is the one which minimizes the loss function\footnote{The subscript $\nabla_{\mathbf{x}}$ is meant to emphasize that the divergence is taken with respect to the data dimension, not the parameters $\theta$.}
\begin{equation}
    \ell_{\rm ISM} = \mathbb{E}_{\mathbf{x} \sim p_t} \left [ \nabla_{\mathbf{x}} \cdot \mathbf{s}^{(t)}_\theta(\mathbf{x}) + \frac{1}{2} ||\mathbf{s}^{(t)}_\theta(\mathbf{x})||^2 \right ].
    \label{eq:ISM}
\end{equation}
This is known as \emph{implicit score matching}~\cite{hyvarinen2005estimation} because nothing about $p_t$ need be known explicitly: all that is required to compute the expectation is samples from $p_t$, which can be obtained from the forward process Eq.~(\ref{eq:VanillaForward}). Several alternative loss functions can be used which rely on the fact that the noise process is Gaussian. In particular, the key insight of~\cite{Ho:2020ddpm} is that all of the conditional distributions in the reverse process may themselves be thought of as Gaussians with learnable means and fixed variances, such that learning the score is equivalent to predicting the mean of the reverse process, which is equivalent to predicting the noise of the forward process. 

\subsection{Review of flow matching models}

Flow matching attempts to learn a time-dependent velocity field $\mathbf{v}_\theta(\x, t)$ that defines an ordinary differential equation (ODE) $\frac{d\x}{dt} = \mathbf{v}_\theta(\x,t)$ whose flow maps a simple prior distribution $\tilde{p}(\x)$ at $t=1$ to the data distribution $p_0(\x)$ at $t=0$. The key idea is that if the conditional paths between paired samples $\x \sim p_0$, $\tilde{\x} \sim \tilde{p}$ are chosen to be straight lines (i.e.  geodesics in Euclidean space), the resulting marginal velocity field approximates the optimal transport map between $p_0(\x)$ and $p(\x)$~\cite{tong2024improving}. In contrast to the stochastic Langevin dynamics of diffusion models, flow matching is invertible and deterministic, enabling exact likelihood computation if the prior can be evaluated. Further, because generation is done through ODE integration, flow matching generally requires fewer function evaluations ($\mathcal{O}(10-100)$ steps) than diffusion models ($\mathcal{O}(1000-10000)$ steps) for often superior generation quality with much less compute.\footnote{See Ref.~\cite{lai2025principlesdiffusionmodels} for a more detailed discussion of the relationship between flow matching and diffusion models, including the relation to optimal transport.}

\subsection{The \texttt{RAMBO} algorithm and $q$-space}
\label{sec:qspace}

In standard applications of generative modeling, no assumptions are made about the data geometry. The intuition that deep generative models can learn distributions on low-dimensional data manifolds embedded in high-dimensional spaces \cite{lai2025principlesdiffusionmodels} is quite appealing, but it is impossible for probability flow to generate a distribution which is, strictly speaking, a set of measure zero in the ambient space. Thus, if a prerequisite for the generative model is that the data manifold constraints be realized exactly, an alternative construction is needed.

Our solution to keeping the probability flow on the phase space manifold is based on the \texttt{RAMBO} construction for sampling massless $N$-particle phase space~\cite{Kleiss:1985gy}, which we now briefly review. Let $Q = \{\mathbf{q}_I \}$ be an \emph{unconstrained} set of $N$ 3-vectors drawn from the distribution
\begin{equation}
\label{eq:pq}
   p_{\rm ref}(\mathbf{q}_1, \cdots, \mathbf{q}_N) \prod_{I=1}^N d^3 \mathbf{q}_I = \prod_{I=1}^N q_I e^{-q_I} \, dq_I \, d\Omega_I,
\end{equation}
where $q_I = |\mathbf{q}_I|$ and $d\Omega_I$ is the angular measure for particle $I$. In other words, the $\mathbf{q}_I$ are isotropically distributed in angle and have magnitudes distributed as $q_I e^{-q_I}$, but the particles are uncorrelated with each other. Construct the massless 4-vectors $q_I^\mu \equiv (q_I, \mathbf{q}_I)$ and apply a conformal transformation $\{q_I^\mu\} \to \{p_I^\mu \}$ consisting of a Lorentz boost along the 3-vector $\mathbf{b} = - \sum_{I=1}^N \q_I$ to the CM frame and a homogeneous rescaling $x$ of all Cartesian momentum components that brings the total energy to unity. Ref.~\cite{Kleiss:1985gy} then shows that the resulting point $P = \{\mathbf{p}_I \}$ will be uniformly distributed on phase space according to the measure Eq.~(\ref{eq:LIPS}) with $\sum_I |\p_I| = 1$ and $\sum_I \p_I = 0$. Since conformal transformations form a group, this construction can be generalized to sample events in any frame with any total energy, simply by boosting and rescaling the resulting distribution.

An intuitive explanation for why \texttt{RAMBO} works comes from interpreting the $e^{-q_I}$ factor in $p_{\rm ref}$ as a Boltzmann weight.\footnote{To the best of our knowledge, this interpretation of \texttt{RAMBO} has not appeared in the literature before.} A collection of massless relativistic particles enjoys Lorentz invariance and scale invariance. Since phase space is a compact manifold with finite volume, the maximal-entropy distribution on phase space is uniform with respect to the Lorentz-invariant volume measure $d\Pi_N$. For a non-compact space, additional constraints must be imposed to render the maximal-entropy distribution normalizable. For example, the Boltzmann distribution is the maximal-entropy distribution on $\mathbb{R}^n$ with a fixed mean energy. \texttt{RAMBO} exploits the fact that it is very easy to sample this maximal-entropy distribution on \emph{unconstrained} products of 1-particle phase space, which is non-compact: particles are uncorrelated, and the energy distributions are simply Boltzmann factors with energies $E_I = |\q_I|$:
\begin{equation}
p_\text{ref}(q^\mu_I,\dotsc,q^\mu_N)\prod_{I=1}^N d^4q_I\, \delta(q_I^2) = \prod_{I=1}^N\frac{d^3{\bf q}_I}{2|{\bf q}_I|}\, e^{-|\q_I|} = \prod_{I=1}^N\frac{d^3{\bf q}_I}{2|{\bf q}_I|}\ e^{-k\cdot q_I}\,.
\end{equation}
In the last equality, we have written the Boltzmann distribution in a Lorentz-invariant way by using an auxiliary 4-vector $k^\mu = (1,0,0,0)$. Now, because both the measure and the probability distribution $p_{\rm ref}$ are Lorentz-invariant, the probability distribution will remain maximal-entropy under \emph{any} Lorentz transformation; in particular, this holds for the set of boosts which bring any point in $q$-space to the CM frame. After such a transformation, the mean energy will no longer be fixed (and indeed, it will be different for every such transformation), but because both $p_{\rm ref}$ and the measure are uniform in the coordinates $\q_I$, a homogeneous change of variables $\q_I \to \q_I/x$ can fix the energy to unity while leaving the entropy unchanged with respect to the rescaled measure. The \texttt{RAMBO} map then transforms the maximal-entropy distribution in $q$-space into the maximal-entropy distribution on $p$-space.\footnote{Note that this interpretation does not work for massive particles; the energy in the Boltzmann factor is exclusively kinetic energy, but the conserved quantity is the total (kinetic + mass) energy. The rescaling step is therefore insufficient to produce the maximally entropic distribution on massive phase space.}

The key feature of the \texttt{RAMBO} construction is trading the energy-momentum constraints, as encoded by the delta functions in Eq.~(\ref{eq:LIPS}), for a specific distribution of $Q = \{\mathbf{q}_I \}$ in Eq.~(\ref{eq:pq}). We exploit this feature to let our generative models live \emph{entirely} in the auxiliary $q$-space. A point $Q$ in $q$-space is parameterized by the boost vector $\mathbf{b}$ required to transform the $N$ 3-vectors to the CM frame, the rescaling $x$ required to obtain unit energy, and the resulting phase space point $P$. The dimension of $q$-space is thus $3 + 1 + (3N-4) = 3N$, the same as the dimension of the ambient space, which makes sense because no constraints are imposed in $q$-space. Since the Lorentz boost coefficients are nonlinear functions of $Q$ (see Ref.~\cite{Kleiss:1985gy} for explicit formulas), the map $Q \to P$ may be viewed as a nonlinear projection onto the phase space manifold. Alternatively, treating $\mathbf{b}$ and $x$ as parameters of the transformation rather than as functions of the coordinates, the map $Q \to P$ is linear and invertible. Thus, given a point $P$ in phase space and parameters $\mathbf{b} \in \mathbb{R}^3$, $x \in \mathbb{R}^+$, we can invert the map to obtain $P \to Q(P, \mathbf{b}, x)$. Applying the \texttt{RAMBO} map to this point $Q$ will of course return the original point $P$, but this interpretation makes clear the relationship between the dimensions of $q$-space and $p$-space, and also provides a way to transform training data from physical phase space to the (unphysical) auxiliary space where the generative model is trained.

\subsection{Diffusion in $q$-space}
\label{sec:qdiffusion}

To train a diffusion model in $q$-space, rather than targeting an isotropic Gaussian distribution, we target the distribution in Eq.~(\ref{eq:pq}), such that the ``pure noise'' distribution, when mapped back to the physical $P$ variables, is the uniform distribution on phase space. This has several advantages:
\begin{itemize}
\item Given a finite-volume Riemannian manifold, the natural endpoint of the forward process is the uniform distribution with respect to the volume measure~\cite{2022arXiv220202763D}.
\item Uniform phase space is easy to sample from, using \texttt{RAMBO}, in order to initialize the reverse process.
\item Any learned deviations from uniform phase space in the trained diffusion model can be straightforwardly interpreted as correlations between particles separate from those induced by energy-momentum conservation. 
\end{itemize}

In our forward process, we target Eq.~(\ref{eq:pq}) by adding a drift term corresponding to its score function. To avoid problems with trying to take a large step in spherical coordinates where we might run past $q_I = 0$, we convert the measure back to Cartesian coordinates and write
\begin{equation}
    p_{\rm ref}(Q) = \prod_{I=1}^N \frac{e^{-q_I}}{q_I} \implies \log p_{\rm ref}(Q) = -\sum_{I=1}^N\Big [q_I + \log (q_I) \Big].
    \label{eq:pref}
\end{equation}
The score is then
\begin{equation}
\label{eq:qscore}
    \nabla \log p_{\rm ref}(Q) = \left \{ -\left(1 + \frac{1}{q_I}\right)\frac{\mathbf{q}_I}{q_I} \right \} \in \mathbb{R}^{3N},
\end{equation}
for every particle $I = 1, \dots N$. Note that this score is manifestly permutation-equivariant, treating each particle on equal footing.

The forward process for our diffusion algorithm is as follows:
\begin{enumerate}
    \item Start from a point $P_0$ in phase space, and map it to a point $Q_0(P_0, \mathbf{b}, x)$ in $q$-space as described in Sec.~\ref{sec:qspace}. There are many possibilities for choosing this transformation $(\mathbf{b},x)$, which we describe further below.
    \item Implement the Langevin dynamics in $q$-space,
    \begin{equation}
        Q_{t+1} = Q_t + \gamma_t \nabla \log p_{\rm ref}(Q_t) + \sqrt{2\gamma_t}Z_t, \qquad t = 0, 1, \dots T,
        \label{eq:qforward}
    \end{equation}
    where $Z_t \sim \mathcal{N}(0,I_{3N \times 3N})$ is isotropic Gaussian noise in $\mathbb{R}^{3N}$, $\gamma_t$ is a fixed noise schedule, and $\nabla \log p_{\rm ref}$ is given in Eq.~(\ref{eq:qscore}).
\end{enumerate}

At any timestep $t$, \texttt{RAMBO} gives a unique, well-defined mapping $Q_{t} \to P_{t}$ where $P_{t}$ lives on $N$-particle phase space, and thus we can map the entire diffusion flow onto physical phase space, either at the level of individual samples, or for the entire distribution. Performing Langevin dynamics in $q$-space has the additional practical advantage that a random rescaling $x$ tends to bring the typical magnitudes $q_I$ to order-1, even if the $p_I$ are quite small, eliminating the need for any explicit data normalization (see Sec.~\ref{sec:Antenna} for an example). 

The $P \to Q$ map may be implemented in multiple ways: for example, 
\begin{itemize}
    \item using a single $(\mathbf{b},x)$ for every point $P$ in the training set, corresponding to diffeomorphically embedding the phase space manifold into $q$-space as a $3N-4$-dimensional submanifold;
    \item using $N_{\rm mult}$ different values of $(\mathbf{b},x)$ for each point $P$, which ``copies'' the phase space submanifold $N_{\rm mult}$ times into $q$-space;
    \item using a \emph{different} $(\mathbf{b},x)$ for each point $P$ to continously ``fill out'' the data distribution in $3N$-dimensional $q$-space by including the four dimensions parameterized by $\mathbf{b}$ and $x$; or
    \item taking the identity map $\mathbf{b} = \boldsymbol{0}, x = 1$ as the embedding.
\end{itemize}
Furthermore, the choice of $(\mathbf{b},x)$ can be deterministic or random; to obtain random $(\mathbf{b}, x)$ following the correct distribution in $q$-space, one may either use the explicit joint distribution of $\mathbf{b}$ and $x$ given in Ref.~\cite{Kleiss:1985gy}, or more practically, sample a random point $\widetilde{Q}$ using \texttt{RAMBO} and extracting the $\mathbf{b}$ and $x$ which would map it to $p$-space. By analogy to the symmetry transformations applied to natural data in AI applications~\cite{wang2025comprehensivesurveydataaugmentation}, we call these transformations ``data augmentation.'' As we will see in the following sections, each data augmentation strategy leads to qualitatively (and quantitatively) different performance, which can be understood geometrically. The data augmentation is a pre-processing step which can be viewed as a hyperparameter of the model.

Once we have obtained a training distribution $p_0(Q)$, as $\gamma_t \to 0$ and $T \to \infty$, the distribution of $Q_t$ tends to $p_{\rm ref}(Q)$, which means that the distribution of $P_t$ tends to uniform on phase space. In practice, we find that achieving statistical uniformity requires a careful choice of the noise schedule $\gamma_t$; in particular, large step sizes tend to underpopulate the ``corners'' of phase space where particles have small energies. There is an additional interplay of the diffusion schedule with the data augmentation procedure, as data which is promoted to $q$-space with the identity map is much farther away from the equilibrium distribution than data which has been continuously augmented; see App.~\ref{app:DiffusionModelDetails} for several examples.

To train a diffusion model, we can learn an additional time-dependent score term $\mathbf{s}^{(t)}_\theta(Q)$ to approximate the true score $\nabla \log p_{t}$ of the distribution of $Q_t \sim p_t(Q)$, such that the reverse process
\begin{equation}
    \bar{Q}_{t+1} = \bar{Q}_t - \bar{\gamma}_t \nabla \log p_{\rm ref}(\bar{Q}_t) + 2 \bar{\gamma}_t \mathbf{s}^{(T-t)}_\theta(\bar{Q}_t) + \sqrt{2\bar{\gamma}_t}Z_t
    \label{eq:reverse}
\end{equation}
with $\bar{Q}_t = Q_{T-t}$ and $\bar{\gamma_t} = \gamma_{T-t-1}$, approaches the desired target distribution $p_0$ as $T \to \infty$. The score function $\mathbf{s}_\theta$ is just a vector in $\mathbb{R}^{3N}$ with no additional geometric constraints; since any point in $\mathbb{R}^{3N}$ is a valid point in $Q$-space, the phase space constraints and measure are \emph{entirely} accounted for by the conformal mapping $Q \to P$. In order to improve the performance of the diffusion models, we borrow a few tricks which have proved useful in image diffusion models: weighting the score-matching loss towards later times in order to obtain a reverse process which evolves at a roughly constant rate from uniform phase space to the target distribution, and using a modified loss which scales better for large $N$ than the divergence $\nabla \cdot \mathbf{s}_\theta$. Details on our architecture and hyperparameter choices are given in Appendix~\ref{app:DiffusionModelDetails}.

Finally, let us remark on the Lorentz transformation properties of our diffusion model setup. We do not impose \emph{anything} about Lorentz invariance at the level of the architecture; indeed, this would seem very difficult to accomplish given that the $q$-space score function $\nabla \log p_{\rm ref}$ is given in terms of the Cartesian components of $\q$, rather than any Lorentz-invariant objects. However, the beauty of diffusion models is that because we are not trying to parameterize a probability distribution analytically, but rather are only interested in generating samples from it, manifestly Lorentz-invariant architectures are not necessary. Indeed, if the samples from the reverse process correctly reproduce the Lorentz-invariant correlations of the training data, then the distribution is \emph{by construction} Lorentz covariant: we can simply boost to any other desired frame. Another way to see this is that because the diffusion process takes place exclusively in $q$-space, we are free to map the entire diffusion flow from $q$-space to any other frame with any other desired total energy using the appropriate variant of the \texttt{RAMBO} map. Lorentz covariance of the distribution is assured because the \texttt{RAMBO} map consists of (invertible) Lorentz and scaling transformations.

\subsection{Flow matching in $q$-space}
\label{sec:qflow}

To implement flow matching in $q$-space, we first map the $p$-space data to $q$-space as described in Sec.~\ref{sec:qspace} and step 1 of our diffusion algorithm in Sec.~\ref{sec:qdiffusion} above. Then, given a training sample $Q$ and a sample $\widetilde{Q}$ drawn from some prior distribution, we define the conditional path
\begin{equation}
    Q_t = (1-t)\, Q + t\, \widetilde{Q}, \qquad t \in [0, 1],
    \label{eq:qflow_path}
\end{equation}
with the conditional velocity field $\mathbf{v} = Q - \widetilde{Q}$. A network $\mathbf{v}_\theta(Q_t, t)$ can be trained to regress this velocity via the loss
\begin{equation}
    \mathcal{L}_{\rm F} = \mathbb{E} \left[ \| \mathbf{v}_\theta(Q_t, t) - \mathbf{v} \|^2 \right].
    \label{eq:qflow_loss}
\end{equation}
To generate new events, we sample $\widetilde{Q} \sim \widetilde{p}(Q)$ and integrate the learned ODE $\frac{dQ}{dt} = \mathbf{v}_\theta(Q, t)$ from $t = 1$ to $t = 0$. The resulting $Q_0$ can then be mapped to physical momenta in $p$-space as before, enforcing energy-momentum conservation. 

Flow matching places no restriction on the prior $\widetilde{p}(Q)$. A natural choice is $p_{\rm ref}(Q)$ in Eq.~(\ref{eq:pref}), which maps to uniform phase space. However, we find empirically that the flow matching network using this prior fails to train. Instead, we find using a fitted Gaussian prior (as in Ref.~\cite{OmniLearned}) in $q$-space to be effective, though unfortunately the Gaussian prior has no direct physical interpretation in $p$-space. While we intend to return to this point in future work, we focus for most of the remainder of this paper on diffusion models, where we are able to incorporate a physically-motivated prior.

\section{Low-dimensional (3-particle) examples}
\label{sec:LowDim}

A prerequisite for applying our construction to unknown data distributions is to verify that we can recover known distributions. In general, determining whether datasets drawn from two high-dimensional distributions are statistically consistent is not an easy problem, due to the curse of dimensionality and the possible presence of arbitrary correlations among the coordinates of the data embedding space. However, massless 3-particle phase space $\Pi_3$ is an excellent proving ground because it is just low-dimensional enough that the restriction to Lorentz-invariant distributions renders it amenable to easy visualization and statistical analysis. $\Pi_3$ is a 5-dimensional manifold locally isomorphic to $\Delta_2 \times {\rm SO}(3)$, where $\Delta_2$ is the Dalitz triangle in the $\mathbb{R}^2$ spanned by the Lorentz-invariant coordinates $s_{12} = 2p_1 \cdot p_2$ and $s_{23} = 2p_2 \cdot p_3$, and the Euler angles of SO(3) orient the ``event plane'' containing $\mathbf{p}_1$, $\mathbf{p}_2$, and $\mathbf{p}_3$.\footnote{The global topology of phase space is that of a sphere~\cite{Batson:2021agz}, but since the diffusion takes place in the topologically trivial non-compact $q$-space, we do not expect topological obstructions to arise even when learning distributions that naturally live on $p$-space.} The uniform distribution on $\Pi_3$ corresponds to a uniform distribution on $\Delta_2$ and uniform sampling from the Haar measure on SO(3). In the CM frame with $p^\mu_{\rm CM} = (1,0,0,0)$, the only possible Lorentz invariants are $s_{12}$, $s_{23}$, $s_{13} = 1 - s_{12} - s_{23}$, and $2p_{\rm CM} \cdot p_1 = 2E_1 = s_{12}+s_{13}$ (and likewise for cyclic permutations of (1,2,3)). Therefore, any deviations from uniform phase space may be visualized as non-uniform distributions in the Dalitz plane and/or anisotropies in the orientation of the event plane.

In this section, we assume that the training data is generated from a weight function $w(\p_1, \p_2, \p_3)$ with respect to the Lorentz-invariant measure, i.e. the probability distribution of the data is
\begin{equation}
 p_0(P) \prod_{I=1}^3 d^3 \p_I \propto   w(\p_1, \p_2, \p_3) \, d\Pi_3  
\end{equation}
where $d\Pi_3$ is given in Eq.~(\ref{eq:LIPS}). In high-energy physics applications, these weight functions come from squared matrix elements computed in QFT, so we will refer to them as matrix elements. Furthermore, because we are always sampling from these distributions and computing score functions rather than normalized probability distributions, we will ignore any normalizing factors and write ``$=$'' rather than ``$\propto$'' when defining the weight functions.

\subsection{Smooth matrix element}
\label{sec:Muon}

\begin{figure*}[!t]
\centering
\includegraphics[width=0.99\textwidth]{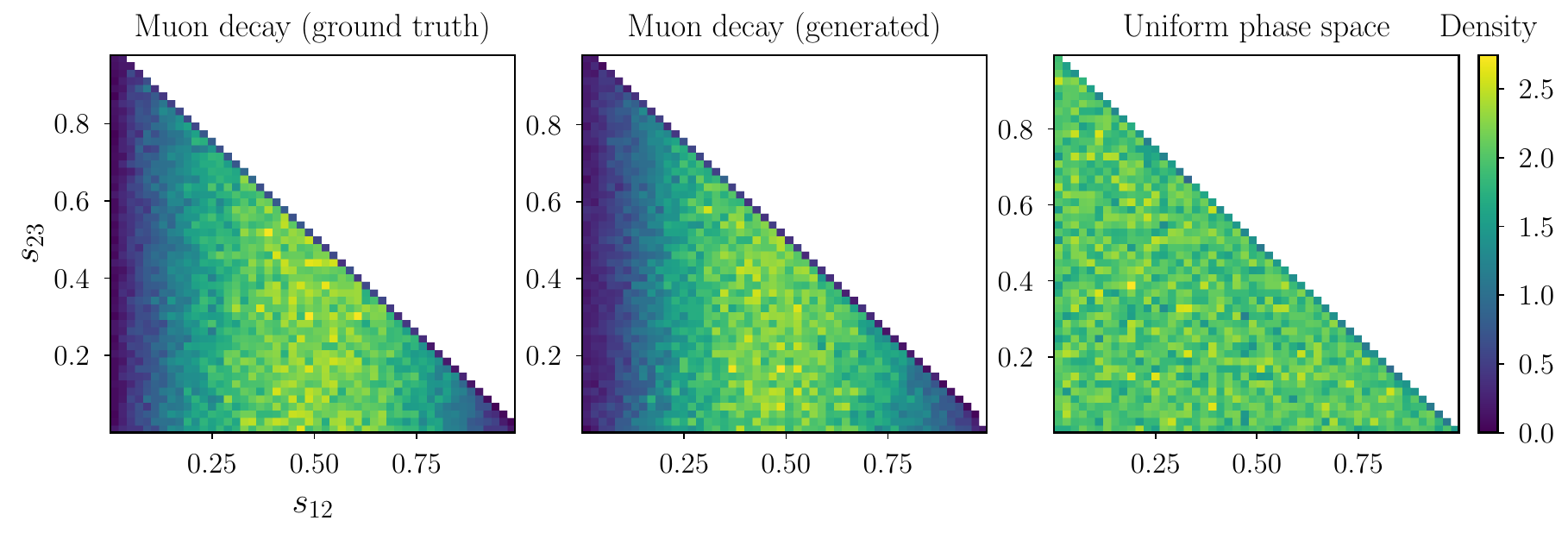}
\caption{(\textit{Left}) Dalitz plot of 100,000 samples from the muon decay distribution~(\ref{eq:w_muon}). (\textit{Center}) Dalitz plot of 100,000 samples from the reverse process of the diffusion model trained on the data in the left panel.  (\textit{Right}) Uniform phase space for comparison.}
\label{fig:muon_Dalitz}
\end{figure*}

A representative example of a smooth Lorentz-invariant weight function on massless $\Pi_3$ is the matrix element for muon decay, $\mu^-\to e^- (p_1) \nu_\mu (p_2) \bar{\nu}_e (p_3)$. Assuming massless decay products and an initial muon 4-momentum $p_{\rm CM}^\mu = (1,0,0,0)$, the distribution on phase space is proportional to
\begin{equation}
    w_{\rm muon} = (p_{\rm CM} \cdot p_3)(p_1 \cdot p_2) = |\p_3|(|\p_1| |\p_2| - \p_1 \cdot \p_2) = \frac{1}{2}|\p_3|(1-2|\p_3|).
    \label{eq:w_muon}
\end{equation}
Machine learning techniques are certainly not needed to learn this simple distribution; muon decay is simply a useful toy example to build intuition about the behavior of our model. A Dalitz-plane histogram of 100,000 samples from this distribution, obtained using rejection sampling, is shown in Fig.~\ref{fig:muon_Dalitz} (left). After training a diffusion model as described in App.~\ref{app:DiffusionModelDetails} according to the algorithm in Sec.~\ref{sec:qspace} with a single $q$-space augmentation $(\mathbf{b},x)$ for each phase space point in the training set, we sample 100,000 times from the trained model to obtain the distribution shown in Fig.~\ref{fig:muon_Dalitz} (center), and compare to the ``pure noise'' distribution of uniform phase space (right). The qualitative agreement is clearly very good.

\begin{figure*}[t]
\centering
\includegraphics[width=0.6\textwidth]{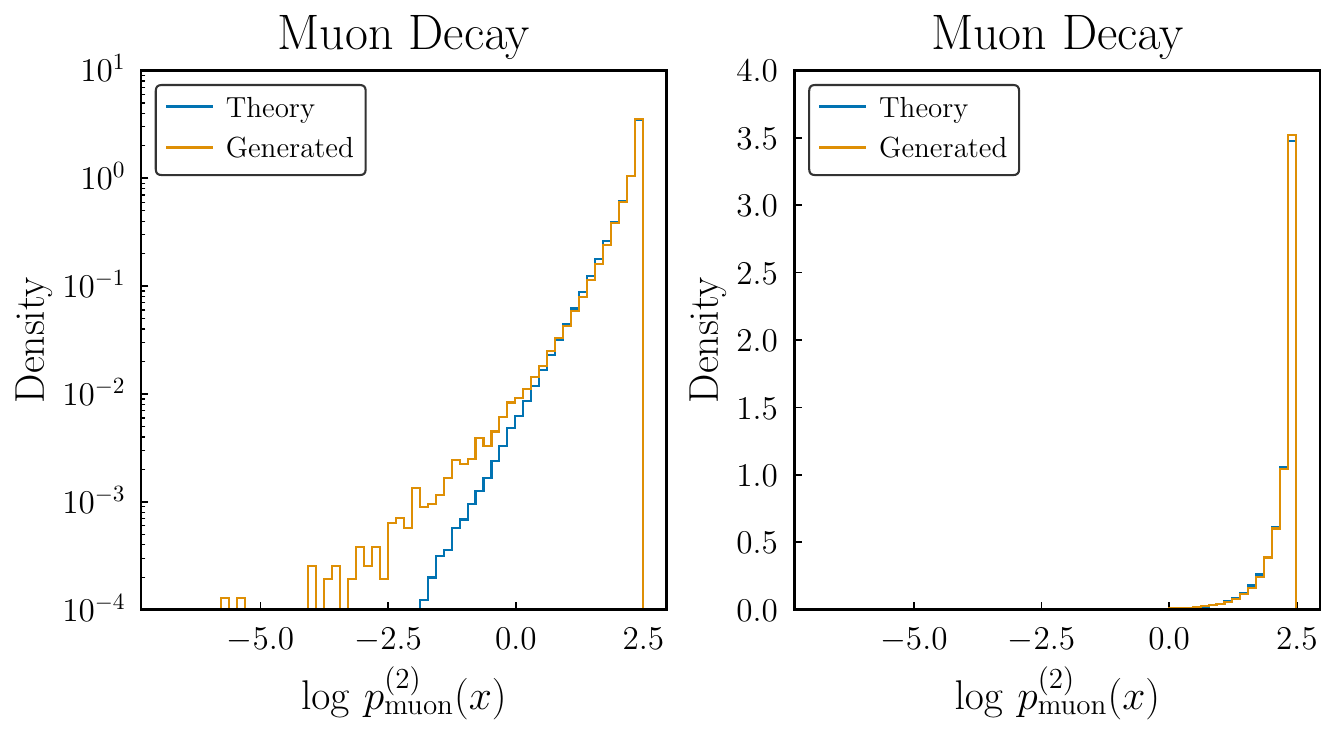}
\caption{
Distributions of the logarithm of the theoretical Dalitz plot PDF for the true muon decay distribution and for our generated distribution.}
\label{fig:Dalitz_LogPDF_KS_Plots}
\end{figure*}

To quantify this agreement in a way that captures some inter-particle correlations, we do the following: Let $p_{\rm muon}^{(2)}(s_{12},s_{23})$ be the probability density function (PDF) of the point in the Dalitz plane corresponding to the phase space point $P = \{\mathbf{p}_I\}$, i.e. the 2-dimensional marginal distribution after integrating Eq.~(\ref{eq:w_muon}) over the three event plane angles. We then evaluate the distribution of $\log(p_{\rm muon}^{(2)})$ for our generated distribution and compare it to the result for the training distribution, given by the matrix element Eq.~(\ref{eq:w_muon}), as shown in Fig.~\ref{fig:Dalitz_LogPDF_KS_Plots}. The agreement is excellent, except on the low-probability tail. A quantitative measure of agreement is given by the Wasserstein-1 (or ``Earth mover's'') distance between the generated and theoretical distributions of $\log(p_{\rm muon}^{(2)})$, which measures the optimal transport distance between distributions; this evaluates to $0.017$. While this value cannot be reliably interpreted in isolation, we compare it to other choices of data augmentation strategy in App.~\ref{app:augmentation}.

\begin{figure*}[t!]
\centering
\includegraphics[width=0.99\textwidth]{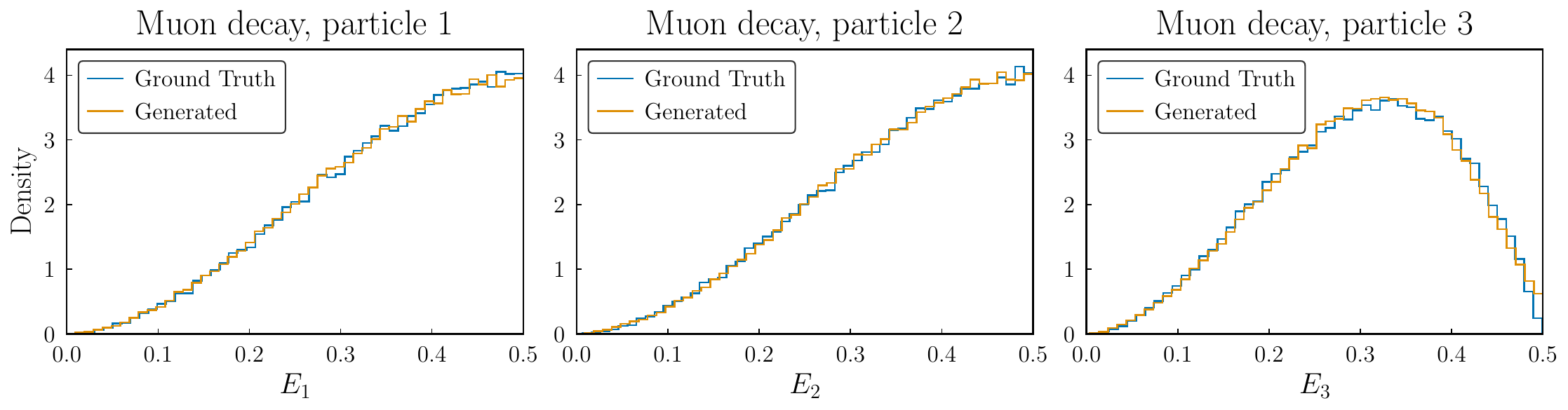}
\caption{Energy distributions for the muon decay matrix element.}
\label{fig:muon_energies}
\end{figure*}

\begin{figure*}[t]
\centering
\includegraphics[width=0.99\textwidth]{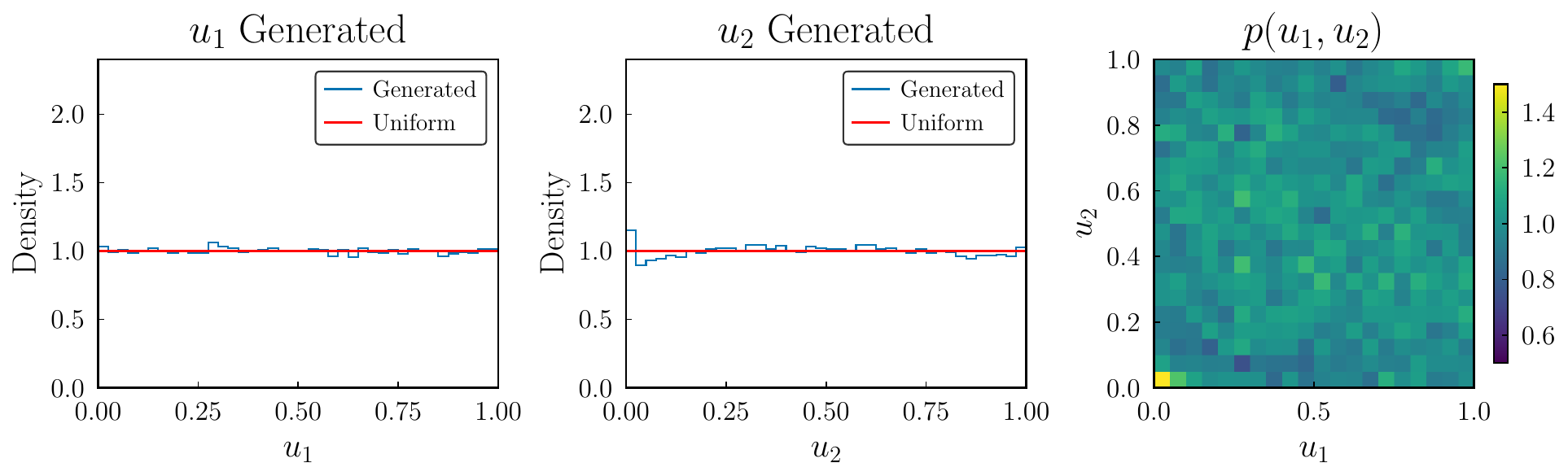}
\caption{Distributions of the Rosenblatt transformation parameters $u_1 = 16 E_1^3 (1-E_1)$ and $u_2 = v^2(3-4v)/E_1^2/(3-4E_1)$ with $v=E_1 + E_2 - 1/2$ for the true muon decay distribution and for our generated distribution. In the true distribution, $u_1$ and $u_2$ should be independently uniformly distributed on $[0,1]$.}
\label{fig:Rosenblatt_Plots}
\end{figure*}

\begin{figure*}[t!]
\centering
\includegraphics[width=0.99\textwidth]{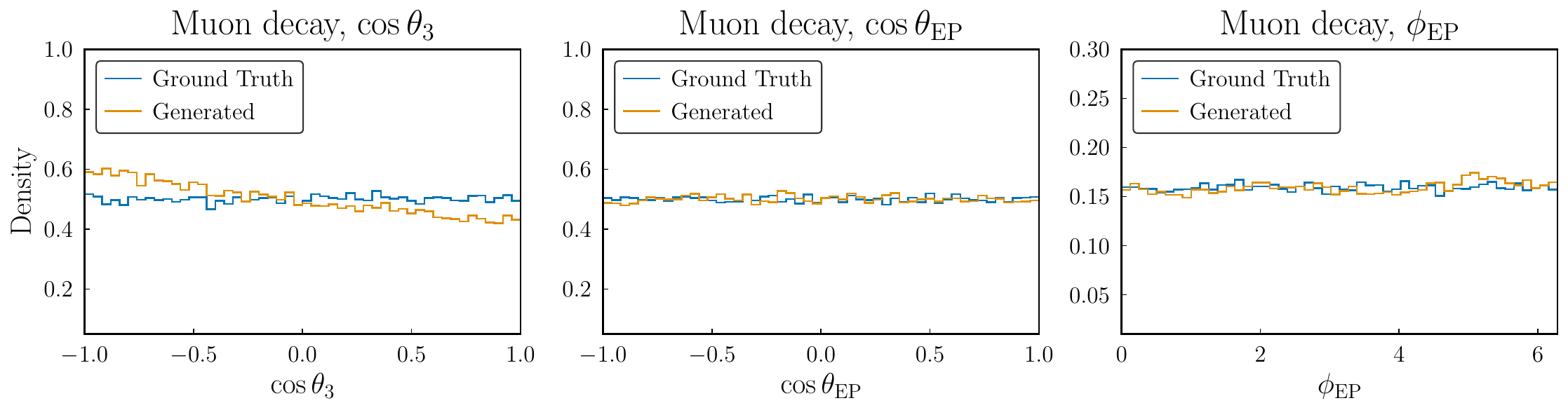}
\caption{Angular distributions for the muon decay matrix element.}
\label{fig:muon_angles}
\end{figure*}

Further confirmation of the accuracy of the learned distribution can be obtained by considering the 1-dimensional marginal distributions of the particle energies, shown in  Fig.~\ref{fig:muon_energies}. Even for the particle 3 energy, where our models tended to oversample near $E_3 \approx 0.5$, the Wasserstein-1 distance is only $0.0016$, indicating good agreement. We additionally employ a Rosenblatt transformation \cite{Rosenblatt}
that maps the theoretical joint 2D distribution of energies $p_{\rm muon}(E_1,E_2)$ to a uniform distribution on $[0,1] \times [0,1]$. Explicitly, we transform to $u_1$ and $u_2$ given by
\begin{equation}
        u_1 = 16 E_1^3 (1-E_1), \qquad 
        u_2 = \frac{v^2(3-4v)}{E_1^2(3-4E_1)} \qquad
        \left(v = E_1 + E_2 - \frac{1}{2}\right).
 \label{eq:RosenblattTransformMuonDecay}
\end{equation}
The Wasserstein-1 distances for the $u_1$ and $u_2$ distributions are $0.0019$ and $0.0030$, respectively, demonstrating good agreement. As shown in Fig.~\ref{fig:Rosenblatt_Plots}, the only obvious discrepancy is in the lowest bin for $u_1$ and $u_2$, corresponding to an overdensity of events at the kinematic endpoint $E_3 = 0.5$.

While the energy distributions show an excellent match, the learned angular distributions (Fig.~\ref{fig:muon_angles}) have a clear residual anisotropy which is not present in the training data.
In particular, although the event plane angles were reproduced well, with Wasserstein-1 distances of only $0.0022$ and $0.019$\footnote{Note that Wasserstein distances scale with variable range, and $\phi_{\rm EP} \in [0, 2\pi)$, rather than $[0,1]$ like most of our variables, so the somewhat larger distance is expected.} for $\cos\theta_{\rm EP}$ and $\phi_{\rm EP}$, respectively, the individual-particle angular distributions are not as well-reproduced, with a Wasserstein-1 distance of $0.060$ for $\cos\theta_3$. In other words, the extrinsic geometry of individual particles is learned incorrectly, but the overall isotropy of the event plane polar angle (and to a lesser extent, the azimuthal angle) is  correctly reproduced. This inconsistency is a direct consequence of the anisotropy induced by picking a single random boost vector $\mathbf{b}$ to augment the data to $q$-space; while $q$-space is unphysical, and every such $\mathbf{b}$ will eventually map back to the same distribution in phase space, the \emph{extrinsic} anisotropy of the $q$-space distribution is being learned along with the \emph{intrinsic} correlations of the desired distribution on phase space. In App.~\ref{app:DiffusionModelDetails}, we show the result of doing two (or more) $q$-space augmentations, as well as continuous augmentation where every training point gets a different $(\mathbf{b},x)$: the single-particle angular distributions improve, but the energy distributions degrade. Similarly, taking the identity map $\mathbf{b} = \mathbf{0}$, $x = 1$ generally gives very poor results for the energy distributions, as the vast majority of the reverse process is spent bringing the $q_I$ in towards the origin in $q$-space, which obscures any correlations in physical phase space.

These considerations illustrate a clear trade-off between learning different aspects of the target distribution, which we do not attempt to optimize in this work, but simply point out as a starting point for future studies. A compact comparison of aggregate metrics across different data augmentation strategies, including additional tests of the overall distributions from trained autoencoders, is presented in App.~\ref{app:augmentation}.

\subsection{Nearly-singular matrix element}
\label{sec:qqg}

\begin{figure*}[!t]
\centering
\includegraphics[width=0.99\textwidth]{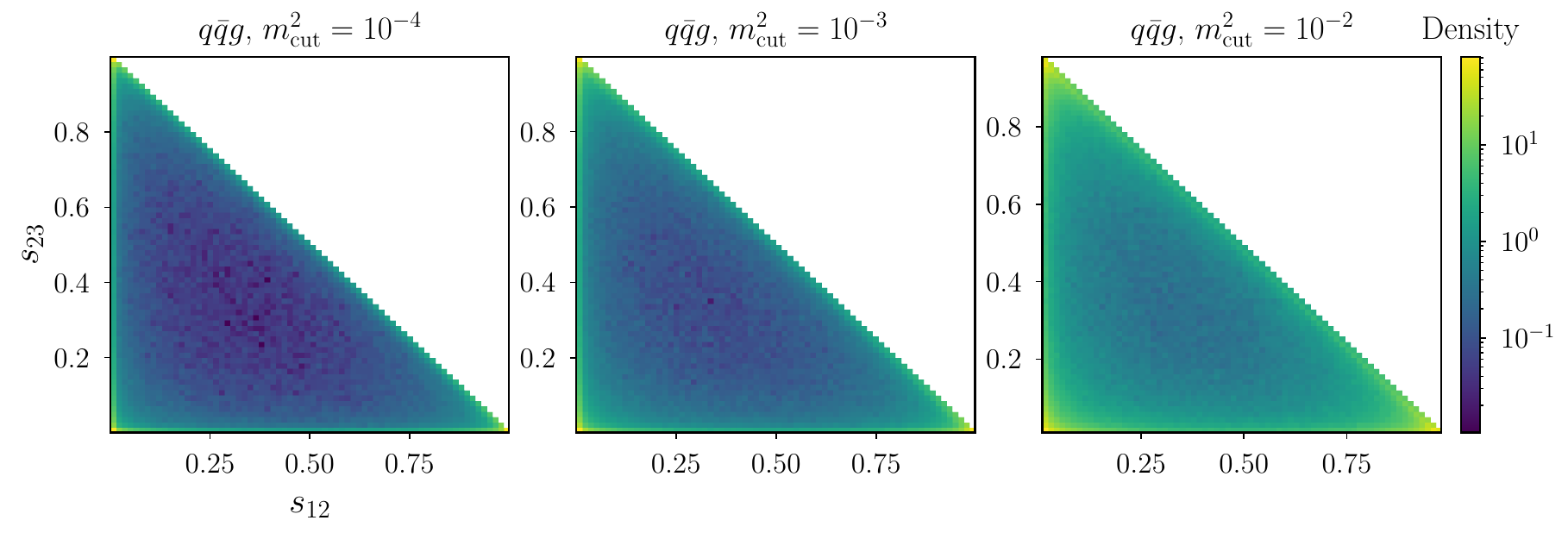}
\caption{Dalitz plots of 500,000 samples from the $e^+ e^- \to q \bar{q} g$ distribution as generated by \texttt{MadGraph}, with varying cutoff $m^2_{\rm cut}$ on the two-particle invariant mass.}
\label{fig:qqg_Dalitz_varycut}
\end{figure*}

In theories with massless gauge fields, fixed-order matrix elements (in other words, distributions on $\Pi_N$ with a fixed number of particles $N$) typically exhibit infrared and collinear singularities. These singularities are a consequence of the fact that standard Feynman diagram perturbation theory is a degenerate perturbation theory. Infrared divergences are resolved by summing inclusively over indistinguishable processes containing arbitrary numbers of particles. This is an interesting and subtle issue for generative models, which typically operate on data with fixed dimension, and one which we will return to in future work. 

For now, though, we study a representative example of a singular distribution on $\Pi_3$, the matrix element for production of a quark/antiquark pair and a gluon radiated off one of the quarks from a color-singlet initial state, $e^+ e^- \to (\gamma^*, Z^*) \to q \bar{q} g$. The matrix element is proportional to
\begin{equation}
 w_{q\bar{q}g} = \frac{(p_q\cdot p_{\bar q}+p_{q}\cdot p_g)^2+(p_q\cdot p_{\bar q}+p_{\bar q}\cdot p_g)^2}{(p_q \cdot p_g)(p_{\bar{q}} \cdot p_g)}\,,
\end{equation}
and thus the distribution blows up when $p_q \cdot p_g \equiv |\p_q| |\p_g| (1 - \cos \theta_{qg}) = 0$, for example. That is, the matrix element diverges when the energy of the gluon vanishes and/or it is collinear to the quark or antiquark. Because the quark has spin-1/2 and couples differently to the photon and the $Z$ boson, the event plane exhibits an anisotropic angular distribution. To render the distribution finite, we generate events in \texttt{MadGraph5\_aMC@NLO 3.5.7} \cite{Alwall:2014hca} with a CM energy of 1 TeV, and impose a minimum invariant mass cutoff, $(p_I + p_J)^2 > m_{\rm cut}^2$. In a real experiment, the quarks and gluons would evolve into jets and hadronize, obscuring their identities, so we symmetrize the events by randomly permuting the order of the final-state particles. To be consistent with our conventions in Sec.~\ref{sec:qspace}, we rescale all 3-vectors homogeneously such that $p_{\rm CM}^\mu = (1,0,0,0)$, and define a dimensionless $m_{\rm cut}^2$ relative to these units (in other words, events generated with $m_{\rm cut} = 10 \ {\rm GeV}$ in \texttt{MadGraph} have $m^2_{\rm cut} = 10^{-4})$.

Dalitz plots of the $q\bar{q}g$ distribution are shown in Fig.~\ref{fig:qqg_Dalitz_varycut}. Physically, points in phase space sampled from $w_{q\bar{q}g}$ consist of two ``hard'' particles which are almost back-to-back and which carry most of the energy, and one ``soft'' particle which is typically low-energy, collinear to one of the hard particles, or both. The distribution thus concentrates on the boundaries and in the corners of the Dalitz plot, with the interior more and more depopulated as $m_{\rm cut}$ is reduced. In contrast to the muon decay matrix element, the symmetrized $q\bar{q}g$ distribution is manifestly permutation-invariant, which is reflected in the $\mathbb{Z}_3$ symmetry of the Dalitz plot.

\begin{figure*}[!t]
\centering
\includegraphics[width=0.99\textwidth]{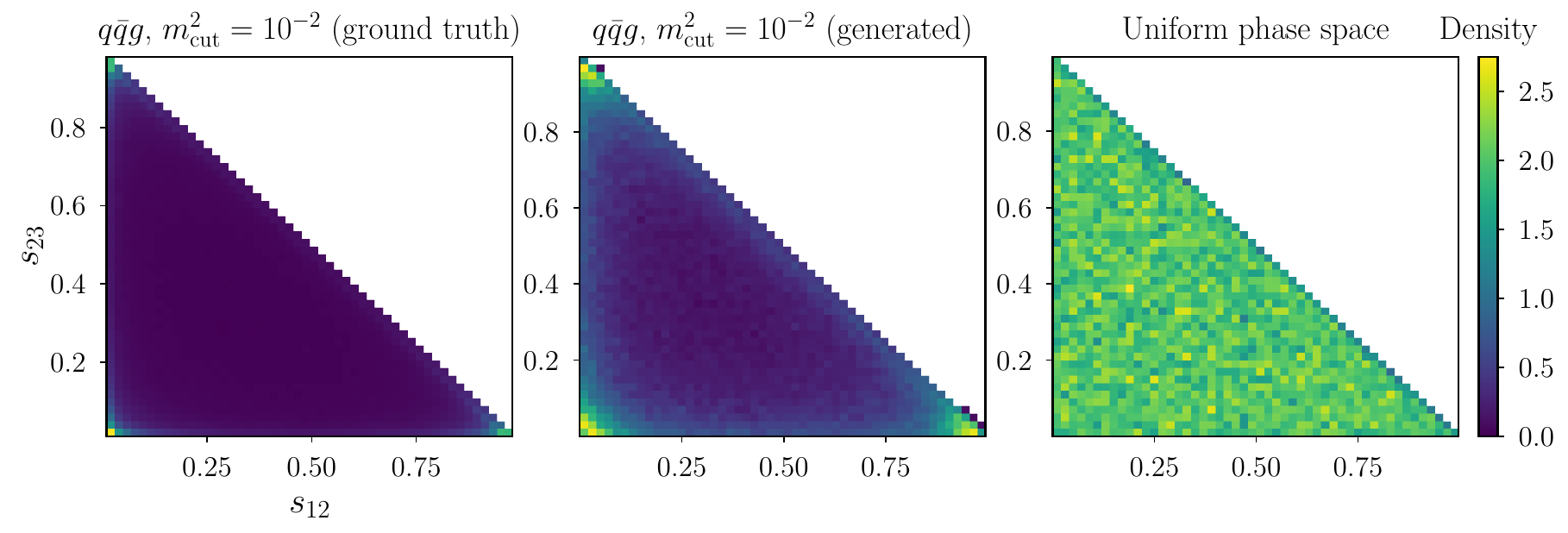}
\caption{As in Fig.~\ref{fig:muon_Dalitz}, but for the $q \bar{q} g$ distribution with $m^2_{\rm cut} = 0.01$.}
\label{fig:qqg_Dalitz_val}
\end{figure*}

\begin{figure*}[!t]
\centering
\includegraphics[width=0.99\textwidth]{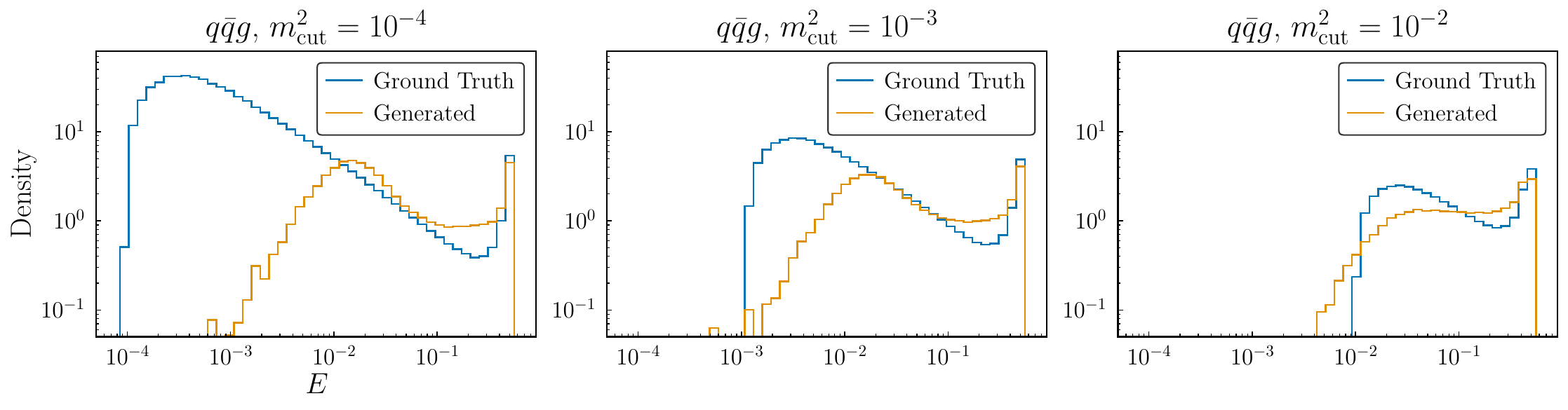}
\caption{Single-particle energy distributions comparing diffusion model samples to the ground truth distribution of $e^+ e^- \to q \bar{q} g$ events for various values of $m^2_{\rm cut}$.}
\label{fig:qqg_energies}
\end{figure*}

\begin{figure*}[!t]
\centering
\includegraphics[width=0.99\textwidth]{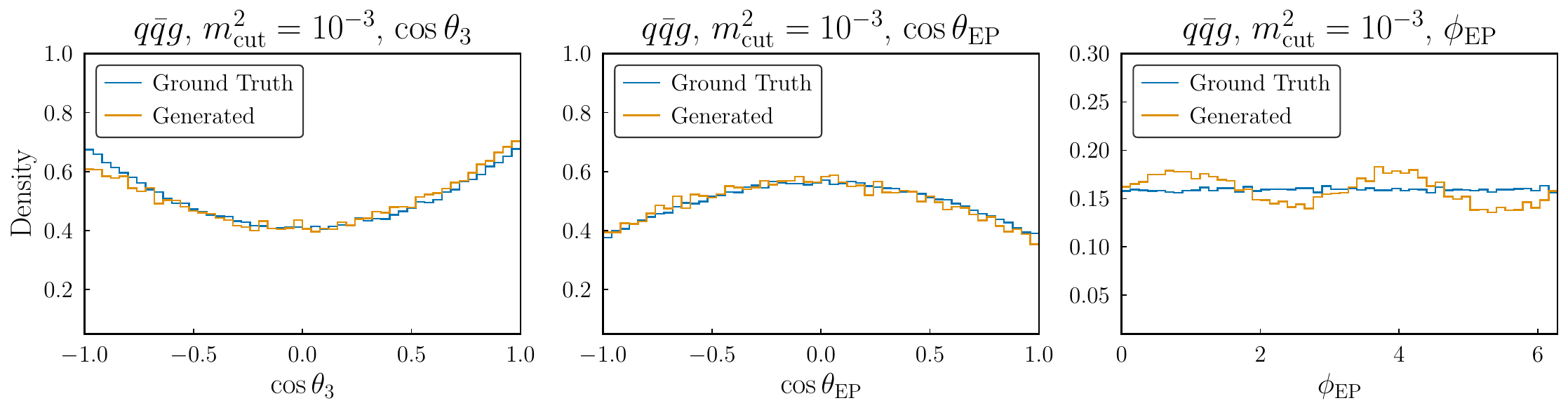}
\caption{Single-particle angular distributions comparing diffusion model samples to the ground truth distribution of $e^+ e^- \to q \bar{q} g$ events for $m^2_{\rm cut} = 10^{-3}$.}
\label{fig:qqg_angular}
\end{figure*}

In Fig.~\ref{fig:qqg_Dalitz_val}, we show the Dalitz plots comparing samples from the trained diffusion model to the ground truth distribution with $m^2_{\rm cut} = 10^{-2}$ and the reference uniform distribution, analogous to Fig.~\ref{fig:muon_Dalitz}. The diffusion model captures the general shape of the distribution well, but the population in the corners is not as sharply peaked in the diffusion samples as it is in the true distribution. This is evident in the single-particle energy distributions, shown in Fig.~\ref{fig:qqg_energies} for various values of $m^2_{\rm cut}$. The diffusion model captures the rough shape of the highest-energy bins, as well as the correct slope at small $E$ for the two smaller $m^2_{\rm cut}$ values, but fails to fully reproduce the low-energy tail. In Fig.~\ref{fig:qqg_angular} we also show the angular distributions for $m^2_{\rm cut} = 10^{-3}$, which demonstrate that the model is learning some of the anisotropy present in the training data, but there are still residual artifacts from the single $q$-space augmentation. As was the case for the smooth distribution in Sec.~\ref{sec:Muon}, higher data augmentation multiplicity leads to better angular distributions but worse energy distributions. For our default choice of 500 diffusion steps, we find that the forward process does not fully converge to the uniform distribution, but this does not seem to be detrimental to quality of the samples from the endpoint of the reverse process -- see App.~\ref{app:DiffusionModelDetails} for more details.

As we have emphasized, though, the low-energy tail of the nearly-singular distribution is not physical, because it depends on the cutoff $m^2_{\rm cut}$. Consider the Lorentz-invariant observable
\begin{equation}
    \tau=\min\{p_I\cdot p_J\},
\end{equation}
where the minimum is taken over all pairs of 4-vectors $I \neq J$ in the event. The observable $\tau$ clearly vanishes when $p_I \to 0$ and when $p_I \cdot p_J \to 0$; any quantity which is insensitive to the presence of an additional soft or collinear particle in the event is known as an IRC-safe observable. However, the distribution of $\tau$ computed on fixed-order matrix elements has an IR divergence due to the presence of soft and collinear singularities. An explicit cutoff $m^2_{\rm cut}$ removes the IR divergence, but renders the distribution of $\tau$ sensitive to the cutoff for values of $\tau$ near zero. The IR divergence is unphysical and is resolved by inclusively summing over events with different numbers of particles; in the case of $e^+ e^- \to q \bar{q} g$, the divergent 1-loop correction to $e^+ e^- \to q \bar{q}$ cancels with the divergence from the additional gluon emission. 

\begin{figure*}[!t]
\centering
\includegraphics[width=0.99\textwidth]{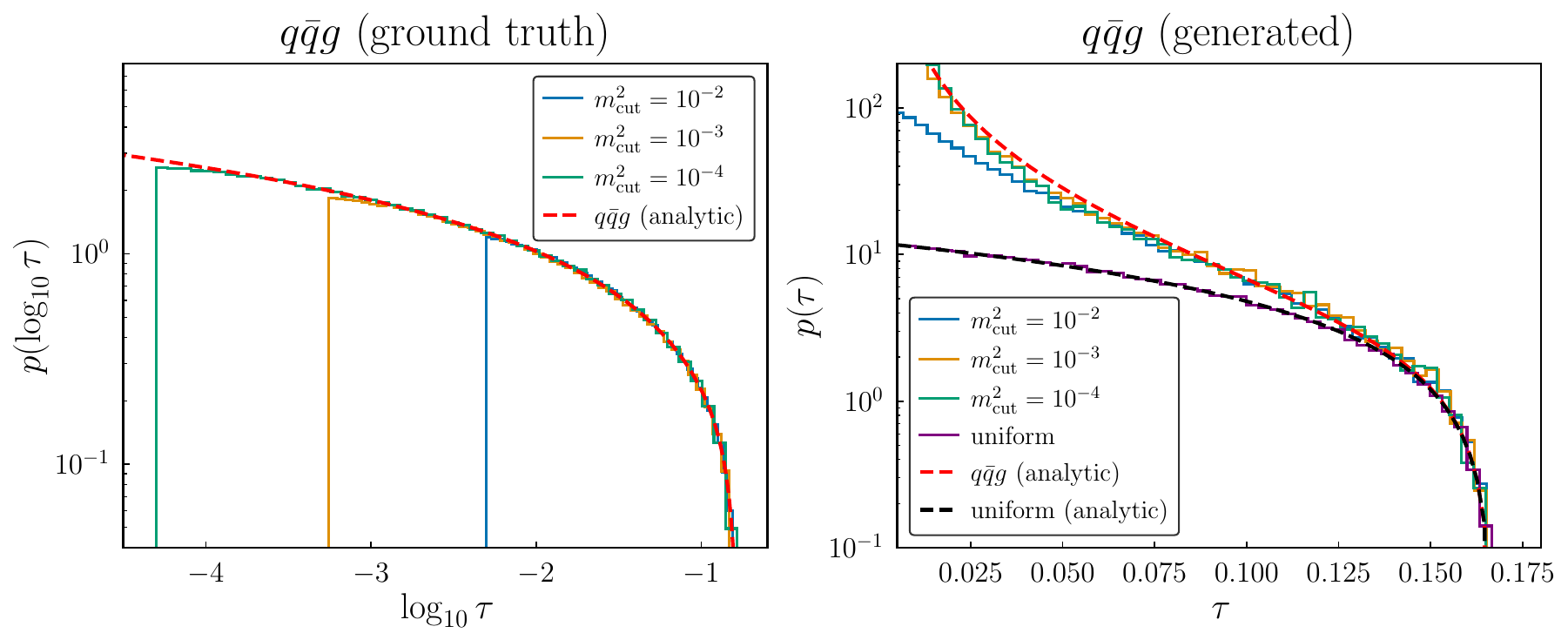}
\caption{(\textit{Left}) Distribution of $\log_{10}\tau$ for 500,000 $q \bar{q} g$ events generated from \texttt{MadGraph}, compared to the analytic prediction in Eq.~(\ref{eq:tauanalytic}). (\textit{Right}) Distribution of $\tau$ for 100,000 samples from diffusion models trained on the $q \bar{q} g$ events from the left panel. For comparison, we also show the distribution for uniform sampling of phase space, along with its analytic prediction in Eq.~(\ref{eq:tauanalytic_uniform3}).}
\label{fig:taudist}
\end{figure*}

The resolution of the singularity only affects the $\tau$ distribution near zero. The values of $\tau$ sufficiently far away from zero are thus independent of the presence of any soft or collinear particles in the event, and the distribution of $\tau$ at its largest values is independent of the cutoff. For the process $e^+ e^- \to q \bar{q} g$, $\tau$ is proportional to thrust \cite{Brandt:1964sa,Farhi:1977sg}, and its distribution is well known \cite{Ellis:1996mzs} and given by 
\begin{equation}
p_{q\bar{q}g}(\tau) \propto 2\frac{6\tau^2-3\tau+1}{\tau(1-2\tau)}\log\left(
\frac{1}{2\tau}-2
\right)-\frac{3}{2\tau}+6+18\tau\,.
\label{eq:tauanalytic}
\end{equation}
In Fig.~\ref{fig:taudist} (left) we plot the distribution of $\log_{10} \tau$ for the ground-truth events. Because the normalization of the histograms is dominated by the IR divergence in the lowest bins, the scale of the $y$-axis is arbitrary, and we rescale each of the histograms in order to demonstrate the excellent agreement of the empirical distribution with the analytical prediction, sufficiently far above the cutoff $\tau_{\rm min} = m^2_{\rm cut}/2.$

In Fig.~\ref{fig:taudist} (right) we plot the distribution of $\tau$ for generated samples from three diffusion models with identical architectures, hyperparameters, and diffusion schedules, trained on the three $q\bar{q}g$ datasets with different $m^2_{\rm cut}.$ The same data augmentation was used for all three models: $\mathbf{b} = (0.3935,  0.4303, -0.3106)$ and $x = 0.087$. We see that the diffusion model distributions are an excellent match to the analytic prediction for $\tau \gtrsim 0.05$, regardless of the cutoff, which corresponds to a dynamic range of nearly 3 orders of magnitude on the $y$-axis.\footnote{The agreement is even better neglecting $m_{\rm cut}^2 = 10^{-2}$, where the training data begins to deviate from the analytic prediction around $\tau = 0.01$.} The learned distribution is also clearly distinct from the $\tau$ distribution for events sampled uniformly from 3-body phase space (see App.~\ref{app:tauder} for a derivation),
\begin{equation}
p^{(3)}_{\rm uniform}(\tau)= 72\left(\frac{1}{6}-\tau\right)\,.
\label{eq:tauanalytic_uniform3}
\end{equation}
That said, our $q$-space construction is guaranteed to correctly match the kinematic endpoint of the distribution at $\tau = 1/6$ which is determined entirely by the geometric constraints of phase space (namely, 3 equally-spaced momentum directions on the sphere $S^2$ separated by angles of $120^\circ$). Remarkably, we see that even though the diffusion model is \emph{not} learning the exact training distribution as a collection of vectors in $\mathbb{R}^9$, it \emph{is} learning the physically-relevant part of the distribution away from the unphysical singular region of phase space. This suggests that $q$-space diffusion could serve as a useful tool for studying distributions of more realistic jets.

\section{High-dimensional example}
\label{sec:Antenna}

At large, fixed particle multiplicity $N$, both analytic matrix elements and Monte Carlo sampling of physical ``hard'' processes (such as $e^+ e^- \to q \bar{q} + 8g$) become extremely unwieldy. Instead, to validate $q$-space generative models on high-dimensional phase space, we will use a toy matrix element that exhibits a similar infrared structure to that of processes in QCD \cite{Parke:1986gb,Berends:1987me}, the ``antenna pole structure'' (APS) matrix element with weight
\begin{equation}
    w^{(N)}_{\rm APS} = \frac{1}{(p_1 \cdot p_2)(p_2 \cdot p_3)\cdots(p_N \cdot p_1)}\,
    \label{eq:APS}
\end{equation}
with respect to $d\Pi_N$. Ref.~\cite{Draggiotis:2000gm} presents an efficient algorithm, called \texttt{SARGE}, for sampling from this distribution with a dimensionless infrared cutoff $\xi_m$ controlling the ratios of Lorentz dot products, such that as $\xi_m \to \infty$, the distribution becomes more and more singular. Larger $\xi_m$ thus tends to concentrate the samples in the corners of phase space. In sampling the APS distribution, we randomly permute the order of particles so that the distribution is manifestly permutation-invariant. Generating $10^6$ events using \texttt{SARGE} takes minutes on modern CPUs.

A point sampled from the distribution~(\ref{eq:APS}) for $N$ particles roughly corresponds to two hard particles each undergoing $N/2$ sequential emissions. In the real QCD parton shower, the multiplicity of perturbative emissions off of the hard particles is approximately Poissonian with mean on the order of 
\begin{equation}
   \bar{N} \simeq \frac{\alpha_s}{\pi}\log^2\frac{Q^2}{\Lambda_\text{QCD}^2},
\end{equation}
where $Q$ is the scale of the hard process and $\Lambda_\text{QCD} \simeq1 \ {\rm GeV}$. For $Q\sim 500 \ {\rm GeV}$, this gives $\bar{N} \simeq 5.4$. It is therefore reasonable to consider $N = 10$ as a representative value for the parton multiplicity of TeV-scale back-to-back jets prior to hadronization.

\begin{figure*}[!t]
\centering
\includegraphics[width=0.99\textwidth]{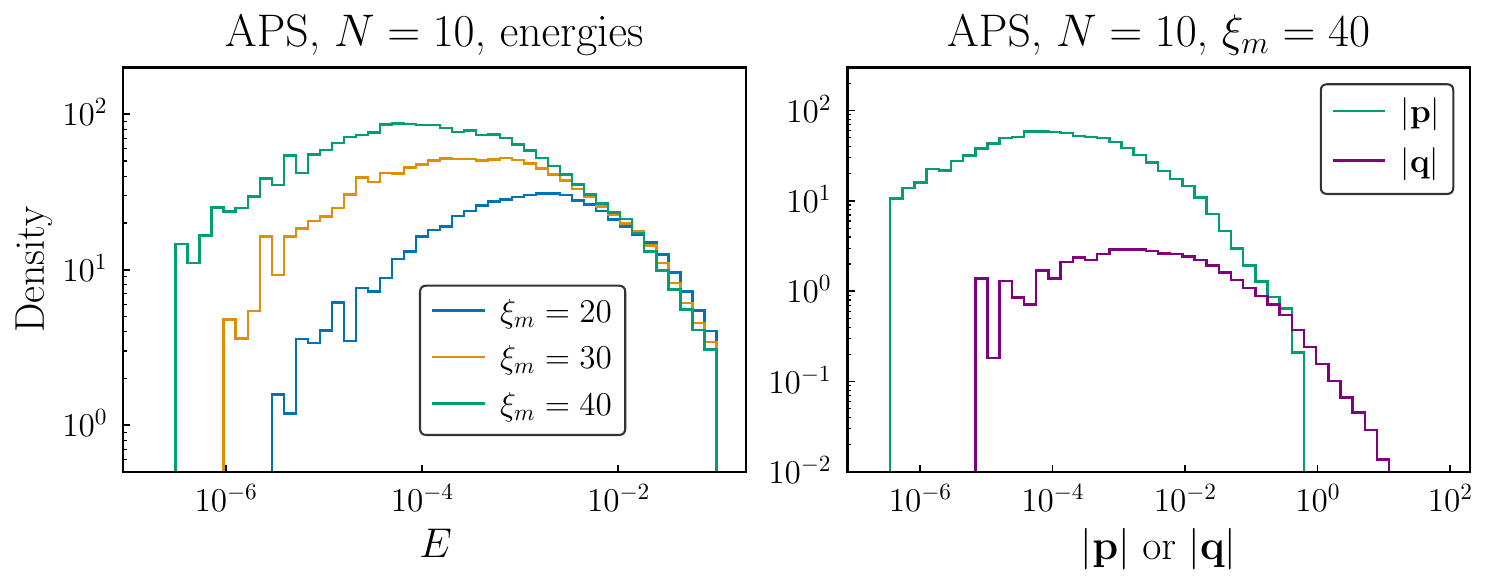}
\caption{(\textit{Left}) Energy distributions for the APS events sampled from Eq.~(\ref{eq:APS}), for various values of the cutoff $\xi_m$. (\textit{Right}) Comparison of $q$-space and $p$-space distributions for $\xi_m = 40$.}
\label{fig:SARGE_training_dists}
\end{figure*}

\subsection{Gaussian diffusion steps for highly-singular distributions}

We generate training data from the \texttt{SARGE} algorithm consisting of $10^{6}$ events with varying cutoffs $\xi_m = 20, 30, 40$. In Fig.~\ref{fig:SARGE_training_dists}, we show the distribution of energies $E_I = |\mathbf{p}_I|$, and for $\xi_m = 40$, we compare the physical energies to the corresponding distribution of $q$-space magnitudes $q_I \equiv |\mathbf{q}_I|$ after performing a random conformal transformation $(\mathbf{b},x)$ on the data. Even though the $q$-space map shifts the magnitudes of the $q$-space vectors to larger values than their phase space counterparts, the highly-singular matrix element results in a large population of $q$-space vectors which are still very close to the origin in $q$-space. If we took this $q$-space distribution as the training distribution and implemented one step in the forward process according to Eq.~(\ref{eq:qforward}), these points would have a score which diverges as $1/q_I$, and would thus be mapped to very large magnitudes even for small step size. On subsequent diffusion steps, these points would receive a relatively small score, and would fail to equilibrate to the target distribution $p_{\rm ref}$ in a reasonable time. 

To address this issue, we implement a Gaussian diffusion phase corresponding to dropping the $q$-space score term and replacing it with a Gaussian score as in Eq.~(\ref{eq:VanillaForward}), for the first $t_{\rm gaus}$ steps. This has the effect of pushing points out from the origin until the $q$-space score has sufficiently small values. For $t_{\rm gaus} \ll T$ and a constant step size $\gamma = 10^{-4}$ for the Gaussian phase, we are able to achieve a convergent forward process; see Appendix~\ref{app:DiffusionModelDetails} for details. For the reverse process, we simply consider the Gaussian and $q$-space score phases as discretizations of a single time-dependent stochastic differential equation with a drift term changing discontinuously at $t = t_{\rm gaus}$, and use the corresponding reverse-process discretization Eq.~(\ref{eq:reverse}) or~(\ref{eq:VanillaReverse}) as appropriate. Improved performance may be achieved by smoothly transitioning between the two phases, but as we will show below, the training is already successful enough to demonstrate the efficacy of our approach that we leave further optimization for future work.

\subsection{$\tau$ distributions for diffusion models}

\begin{figure*}[!t]
\centering
\includegraphics[width=0.99\textwidth]{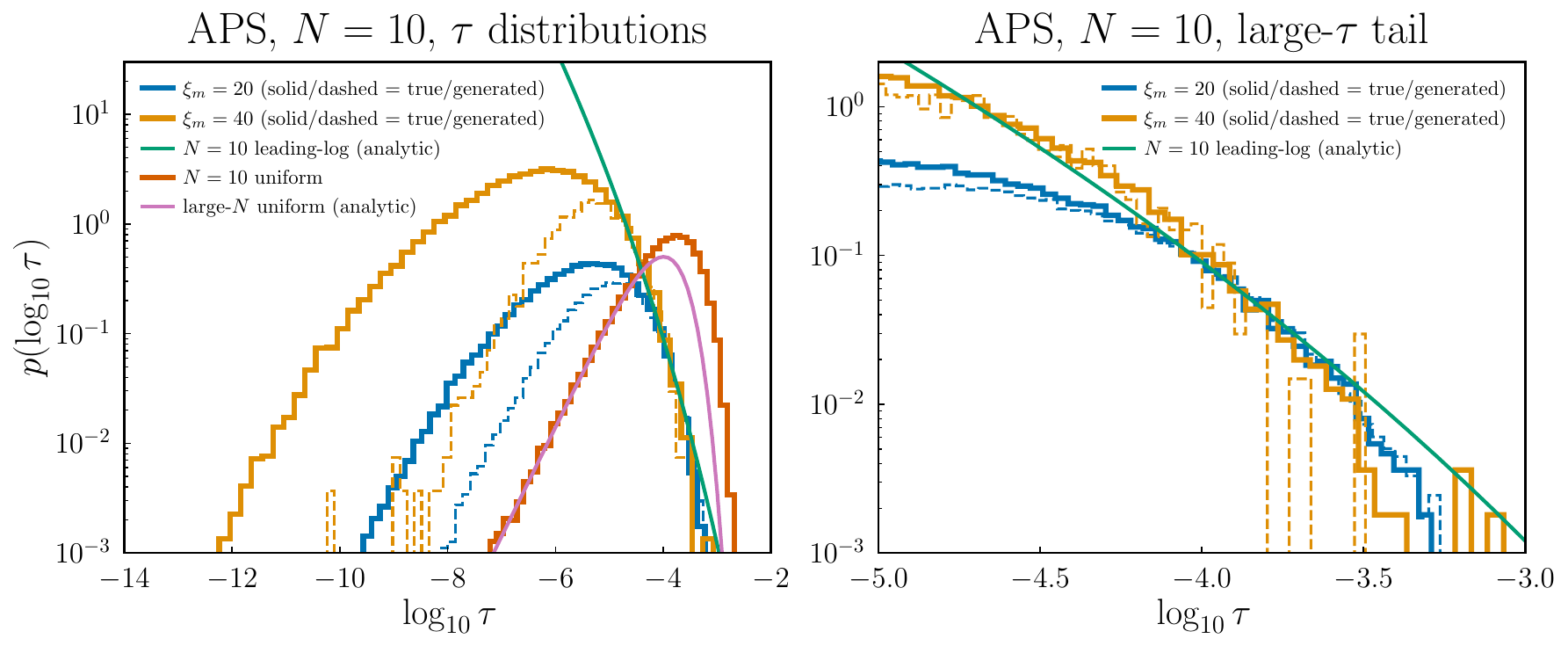}
\caption{Distributions of $\tau = {\rm min}\{ p_I \cdot p_J \}$ for ground-truth APS events (solid) and samples from trained diffusion models (dashed). For comparison, uniform events on 10-particle phase space and the analytic distributions Eqs.~(\ref{eq:pN_APS})--(\ref{eq:pN_uniform}) are also shown. A zoom-in of the large-$\tau$ region is shown at right.}
\label{fig:SARGE_taudist}
\end{figure*}

Fig.~\ref{fig:SARGE_taudist} shows the distributions $p(\tau)$ for generated APS events from diffusion models trained using the same data-augmentation conformal transformation with $\mathbf{b} = (-0.0732,  0.2464, -0.1534)$ and $x = 0.0846$.  We compare to both the ground-truth events and uniform phase space, where the analytic distributions for arbitrary $N \gg 1$ are
\begin{align}
\label{eq:pN_APS}
p^{(N)}_{\rm APS}(\tau) &\propto - \frac{\log^{2N-5}(\tau)}{\tau}\,,\\
\label{eq:pN_uniform}
p_{\rm uniform}^{(N)}(\tau) &= 2N^5\sqrt{2\tau}\,K_1\left(
2N\sqrt{2\tau}
\right)\,\exp\left[
-\frac{N^2}{2}\left(
1-(2N)^2 \tau\,K_2(
2N\sqrt{2\tau})
\right)
\right]\,.
\end{align}
Here, $K_1(x)$ and $K_2(x)$ are modified Bessel functions.  The distribution of $\tau$ on the APS matrix element is the leading-logarithmic contribution, corresponding to $N-2$ strongly-ordered emissions; see, e.g., Ref.~\cite{Larkoski:2019nwj}.  The distribution of $\tau$ on uniform phase space follows from thermalization in the large-$N$ limit.  We present details of the uniform phase space calculation in App.~\ref{app:tauder}.  In contrast to the 3-particle case, reaching the kinematic endpoint of the $\tau$ distribution is exponentially unlikely because it requires all 10 particles to be exactly equally spaced on the sphere. Nonetheless, it is clear that the learned distributions are strongly distinct from uniform phase space, and follow the $\log^{15}(\tau)$ distribution of the ground-truth events at large $\tau$. As was the case for the 3-particle distributions, the diffusion model does not correctly capture small-$\tau$ tail, but this is the unphysical cutoff-dependent part of the distribution; while the cutoff dependence was fairly mild for the single-particle energy distributions in Fig.~\ref{fig:SARGE_training_dists}, it is much more obvious in the $\tau$ distributions, where the three training sets have tails separated by orders of magnitude in $\tau$.

\subsection{Comparing $p$-space and $q$-space generative models}
\label{sec:compareflow}

\begin{figure*}[!t]
\centering
\includegraphics[width=0.7\textwidth]{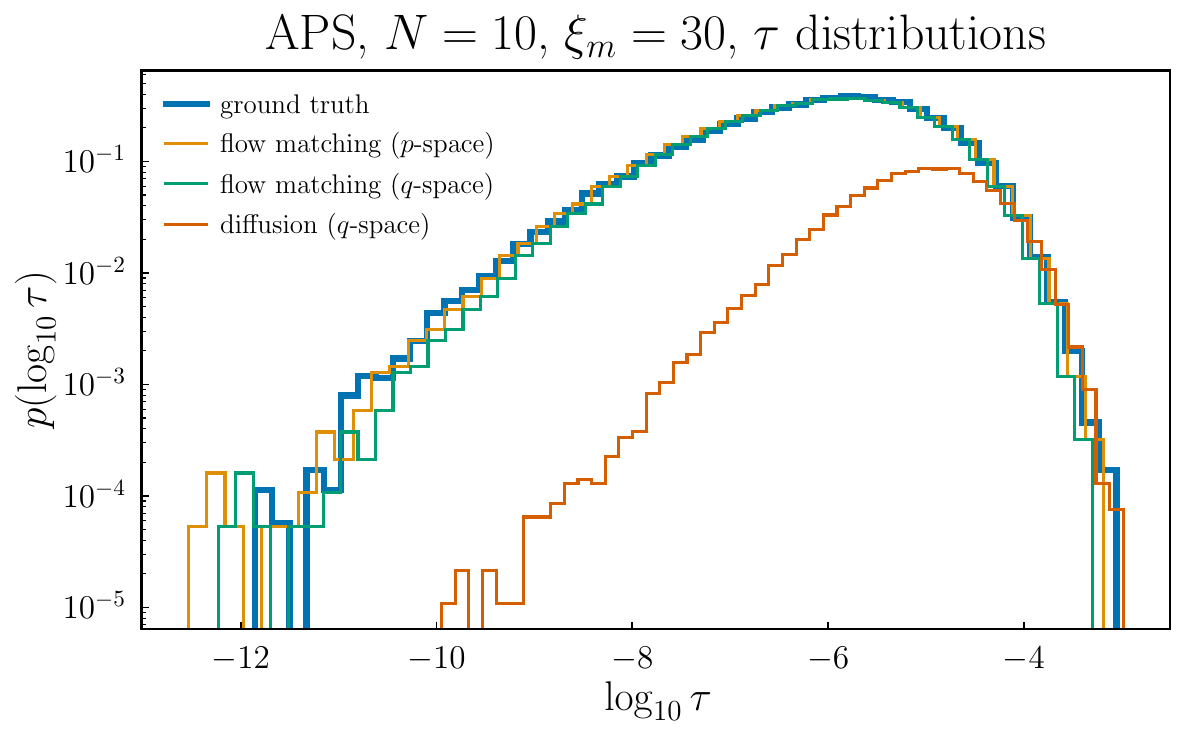}
\caption{Comparison of flow matching models in $p$-space and $q$-space with a $q$-space diffusion model, for the $N = 10$ APS distribution with $\xi_m = 30$.}
\label{fig:tau_dist_q_v_p}
\end{figure*}

To compare our $q$-space diffusion model to existing methods, we train three additional models: Gaussian diffusion and flow matching directly on the 3-momentum vectors $P = \{\mathbf{p}_I\}$, and flow matching in $q$-space as described in Sec.~\ref{sec:qflow}. For all, we use the same \textsc{Point-Edge-Transformer} (PET) architecture and training hyperparameters, see Appendix \ref{app:high-dim-qspace}. Flow matching on $q$-space and $p$-space proceeds as described in Section \ref{sec:qflow} and in \textsc{OmniLearn}~\cite{OmniLearned}. The procedure for diffusion models using Langevin dynamics in both $p$-space and $q$-space is also given in Appendix \ref{app:high-dim-qspace}.

\begin{figure*}[!t]
\centering
\includegraphics[width=0.99\textwidth]{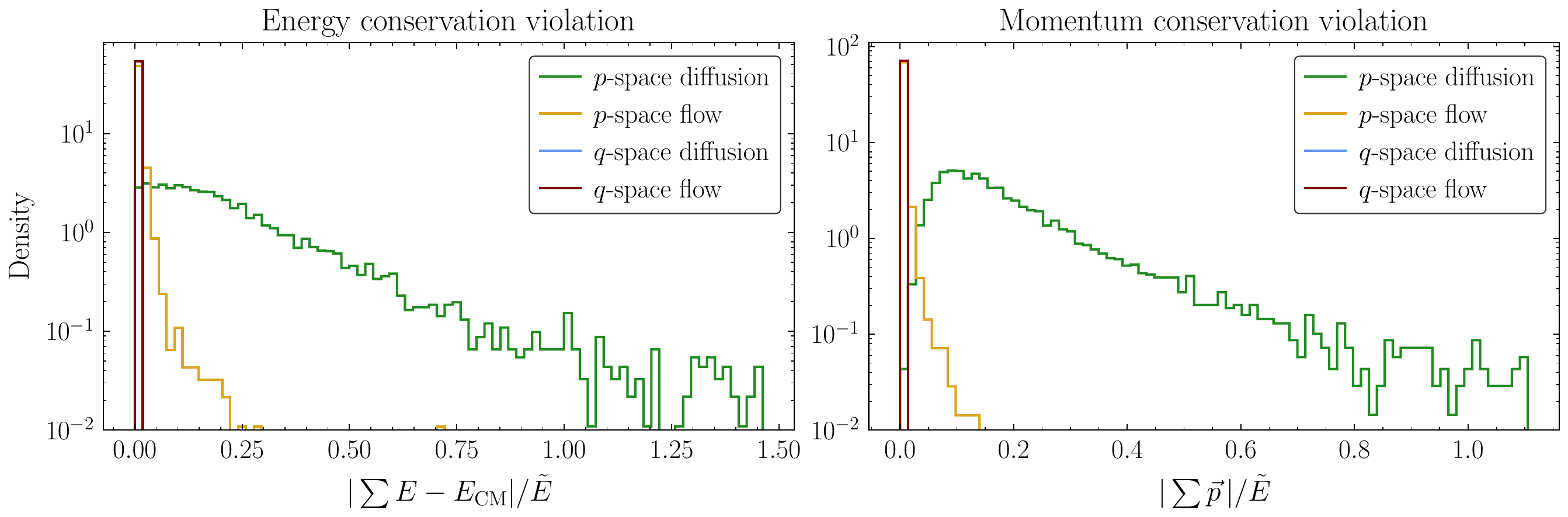}
\caption{Per-event energy and momentum violation normalized by the median single-particle energy $\tilde{E}$.}
\label{fig:conservation_energy}
\end{figure*}

 As with the $q$-space models, $p$-space models learn the shape of physical distributions with high fidelity. Histograms of $\tau = {\rm min}\{ p_I \cdot p_J \}$ are excellent matches to the truth distribution as seen in Figure \ref{fig:tau_dist_q_v_p}, and in particular the low-energy tail is reproduced much better with the flow matching models than with the diffusion model. However, the generated events in $p$-space exhibit substantial violations of energy-momentum conservation, as shown in Figure \ref{fig:conservation_energy} and quantified in Table \ref{tab:conservation}. By contrast, the $q$-space flow matching model does not appear to suffer from reduced accuracy in the reconstructed $\tau$ distribution, matching the ground-truth distribution over an impressive 9 orders of magnitude in $\tau$, while also conserving energy and momentum by construction. Furthermore, we observe improved generation quality and substantially faster sampling with both flow matching models compared to diffusion models. This motivates further study of $q$-space flow matching as a promising generative model for distributions of relativistic particles.

\begin{table}[t]
\setlength{\tabcolsep}{12pt}
    \centering
    \begin{tabular}{lcccc}
    \vspace{0.1cm}
        & $|\sum_I E_I - E_{\rm CM}|/\tilde{E}$ & $|\sum_I p_{I,x}|/\tilde{E}$ & $|\sum_I p_{I,y}|/\tilde{E}$ & $|\sum_I p_{I,z}|/\tilde{E}$ \\
        \hline
        $p$-space diffusion          & $ 0.2585$ & $ 0.1264$ & $0.1220$ & $0.0820$ \\
        $p$-space flow matching          & $0.0094$ & $0.0023$ & $0.0024$ & $0.0023$ \\
        \hline
    \end{tabular}
        \caption{Mean conservation violations for generated events ($N = 10$ particles, $\xi_m = 30$), expressed as fractions of the median single-particle energy $\tilde{E}$ in each event. Generative models in $q$-space satisfy conservation to machine precision (as does the ground-truth data), while generative models in $p$-space exhibit energy and momentum conservation violations that are comparable to the median particle energy.}
            \label{tab:conservation}
\end{table}

\section{Conclusions}
\label{sec:Conclusions}

\subsection{AI for physics}

In this paper we have introduced generative modeling in $q$-space, a framework which exactly implements energy-momentum conservation constraints for distributions of relativistic particles. Lorentz symmetry, rather than being hard-coded into the architecture or learned directly from the data, is incorporated implicitly through the \texttt{RAMBO} map between physical phase space and auxiliary $q$-space. We have shown that diffusion models in $q$-space with flat phase-space priors can learn smooth distributions and the physical parts of nearly-singular distributions that are independent of arbitrary cutoffs, and furthermore that flow matching in $q$-space with (unphysical) Gaussian priors can reproduce the full training distribution. We have not attempted to extensively optimize our setup, but we have illustrated numerous tradeoffs in performance as a function of data augmentation strategy, diffusion schedule, and architecture, many of which have a straightforward geometric interpretation in $q$-space. 

While we focus in this paper on fairly low-dimensional distributions where exact functional forms are known and all of our claimed structure can be validated precisely, our eventual goal is to train these generative models on jet data, with $\mathcal{O}(200)$ particles per event. There are many examples of applying deep generative modeling to jets~\cite{Leigh:2023toe,Mikuni:2023dvk,Butter:2023fov,Leigh:2023zle,Buhmann:2023zgc,Sengupta:2023vtm,Quetant:2024ftg,Jiang:2025pil,Mikuni:2025tar,Dreyer:2025zhp,OmniLearned}, both real and simulated. To our knowledge, none of them take the geometric structure of phase space as a starting point. Our $q$-space construction is a simple drop-in replacement for any generative model operating directly on the 3-momentum components, which can implement exact energy-momentum conservation at no cost in performance, as demonstrated in Sec.~\ref{sec:compareflow}.

\subsection{Physics for AI}

Our focus on diffusion models stems largely from their use as a powerful diagnostic tool for understanding data with hierarchical structure, which has implications far beyond particle physics. Generative AI must be trustworthy in order to be useful. If models cannot be trusted to learn the true semantic or latent structure encoded in their training data, they cannot be used to reliably discover new, unseen patterns, limiting their ability to generalize beyond the distributions on which they were trained. In particular, if one cannot verify that a network has recovered structures already known to be true, it is difficult to trust it as a guide to additional structure \cite{Roscher2020Explainable,Krenn2022ScientificUnderstanding}. This point is even more critical in scientific generative AI, where, on the one hand, empirical verification is relatively straightforward via simulations or real-world experiments, but, on the other hand, the data distribution can be continuously deformed by a set of latent control parameters (for example, coupling constants in the case of distributions derived from QFT) that the models must learn in order to be reliable. Ensuring that generative models learn the underlying structure of a high-dimensional distribution is therefore a difficult task, and failures to generalize across two simulations of the same underlying physics have already been demonstrated~\cite{Gambhir:2025xim}.

Recent work has begun to explore how deep and diffusion-based models learn latent structures such as hierarchical, compositional, and rule-based features in data \cite{Cagnetta2024RandomHierarchy,Sclocchi2025PhaseTransition,Sclocchi2025LatentHierarchy,Favero2025CompositionalGeneralization,Han2025HiddenRules}. However, these studies typically either use synthetic datasets with prescribed hidden rules, or else natural data such as images and text for which the underlying constraints likely exist~\cite{levi2025the,levi2023universalstatisticalstructurescaling} but are not known a priori. Particle physics data is ideally positioned between these two extremes: the distributions are high-dimensional and highly structured, but they are constrained by exact symmetries and conservation laws that must be reflected in the data. Many of the distinguishing features of perturbative QCD that are imprinted on jets -- resolution of soft and collinear singularities in matrix elements by resummation, IRC safety, and parton-level permutation invariance, for example -- descend from the Poincar\'{e} invariance of the underlying theory. While the phase space dimension of an $N$-particle jet is $3N-4$, which is the same order of magnitude as the embedding dimension $3N$, the equivalence class of jets related by IRC-safety can have an intrinsic dimension that is \emph{much} smaller, stemming from the hierarchical structure of jets originating from a few hard partons~\cite{Komiske:2019jim,Batson:2021agz}. This rich structure makes jets an appealing model of data, with which one can hope to understand the success of deep learning on other natural data. That said, the rigid scaffold of the data cuts both ways. Hard-coding too much physics directly into the architecture risks obscuring what structure the network is actually learning and limiting scalability~\cite{Breso-Pla:2026tlz}, while asking an unconstrained model to recover exact continuous conservation laws from finite data is unnecessarily difficult, as we have seen in Fig.~\ref{fig:conservation_energy}.

In this work, we take a first step toward addressing this tension, ensuring deep generative models reliably learn from particle physics data by modifying the generative process rather than the NN architecture itself, so that the sampling trajectory remains exactly on the phase-space manifold while the network remains free to learn the nontrivial correlations in the data. This makes the present construction a controlled setting in which the reverse process can be used as a diagnostic tool, and in which learned correlations can be compared directly to known physical structure, without putting all of that structure into the architecture by hand. In the case of diffusion models, the reverse trajectory is (in principle) directly interpretable because the diffusion timescale can be associated with a changing noise level which progressively reveals different layers of the hierarchical data~\cite{Favero2025CompositionalGeneralization}. In future work, it would be interesting to see if the same strategy can be applied to flow matching; when the prior distribution is that of uniform phase space, the flow velocity field can be assigned a physical interpretation as well.

\subsection{Outlook}

Regardless of the motivation, the choice of generative model, or the performance of the model on various metrics or benchmarks, we emphasize strongly that imposing physics priors will make models trained on physics data \emph{more interpretable}. Our $q$-space framework is a simple module which can simultaneously improve the physical reliability of existing generative models in high-energy physics -- and thus their utility in extracting physical consequences from data -- while also exploiting the structure of high-energy physics data to shed light on AI trained on natural data. We are optimistic that this ``physics for AI for physics'' approach still has much to offer both the high-energy physics and AI communities.

\acknowledgments

We thank Yohai Bar-Sinai,  Alessandro Favero, Josh Foster, Rikab Gambhir, Jesse Thaler, and Jacob Zavatone-Veth for useful discussions, and Gaia Grosso for collaboration in the early stages of this work. I.E. thanks Vinicius Mikuni, Joschka Birk, Wahid Bhimji, and Benjamin Nachman for helpful insight into Point-Edge Transformers. This research is based on work supported by the Canadian AI Safety Institute Research Program at CIFAR. Resources used in preparing this research were provided, in part, by the Province of Ontario, the Government of Canada through CIFAR, and companies sponsoring the Vector Institute. Y.K. and I.E. acknowledge the support of a Discovery Grant from the Natural Sciences and Engineering Research Council of Canada (NSERC). I.E is supported in part by the Connaught International Scholarship at University of Toronto. Y.K. is grateful for the hospitality of Perimeter Institute where part of this work was carried out. Research at Perimeter Institute is supported in part by the Government of Canada through the Department of Innovation, Science and Economic Development Canada and by the Province of Ontario through the Ministry of Colleges, Universities, Research Excellence and Security. Y.K. and I.E. thank Marat Freytsis and Anthropic for access to Claude Code which assisted with aspects of training the generative models used in this work. Y.K. and N.L thank the Aspen Center for Physics, which is supported by National Science Foundation grant PHY-2210452, where portions of this work were completed.  This manuscript has been authored by Fermi Forward Discovery Group, LLC under Contract No.\ 89243024CSC000002 with the U.S. Department of Energy, Office of Science, Office of High Energy Physics. 

\appendix

\section{Architecture and hyperparameter choices for numerical experiments}
\label{app:DiffusionModelDetails}

\subsection{3-particle distributions}
Our diffusion model score network consists of a 4-layer multilayer perceptron (MLP) with \texttt{SiLU} activations and hidden width 256, using default \texttt{PyTorch} initializations for the inner layers but initializing the last-layer parameters to zero. Diffusion time is encoded through sinusoidal embeddings with dimension 64, appended to the 9-dimensional input for 3-particle $q$-space. Training is performed using the \texttt{AdamW} optimizer with a learning rate of $10^{-3}$, weight decay of $10^{-4}$, gradient clipping of 1.0, and a cosine annealing learning rate schedule. We train for 100 epochs on a training set size of 500,000, with batch size 1024, and take the network with the lowest loss over the 100 epochs as our trained network. The default diffusion schedule is a linear schedule going from step sizes of 0.002 to 0.01 over 500 steps. We provide all of these details for the sake of reproducibility; we do not expect that any of these rather arbitrary choices are crucial for the success of the model. We use an implicit score-matching loss function~(\ref{eq:ISM}) weighted by $(1-t/T)$ and oversample at earlier times with a weight of $(1-t/T)^2$, to encourage the network to learn the score at earlier times when the data is less corrupted by noise.

\subsubsection{Data augmentation}
\label{app:augmentation}

\begin{figure*}[t]
\centering
\includegraphics[width=0.99\textwidth]{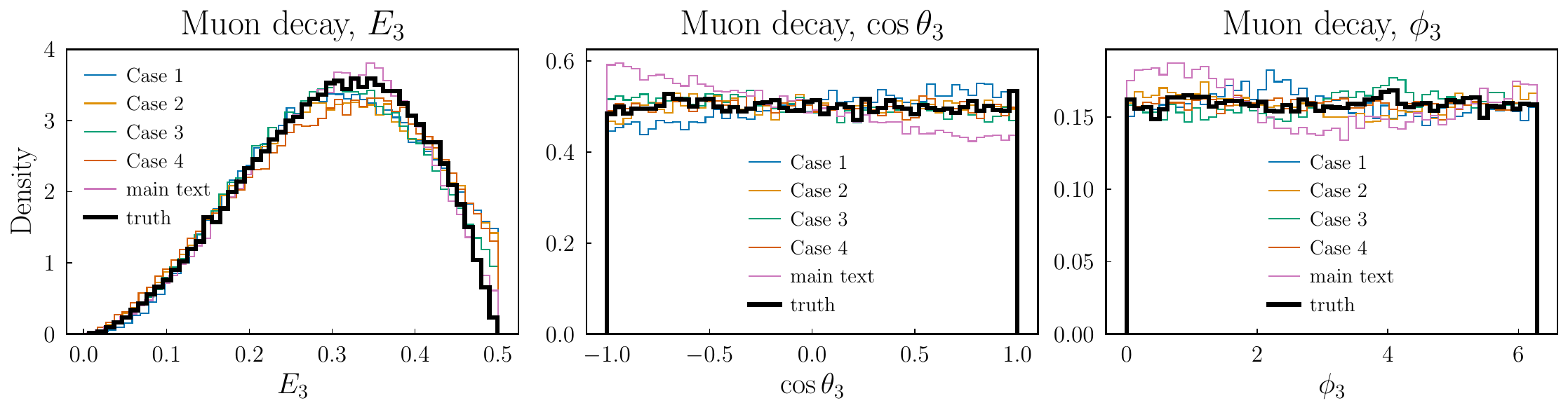}
\caption{Energy (left) and angular (center and right) distributions of particle 3 for varying data augmentation strategies.}
\label{fig:vary_Nmult}
\end{figure*}

\begin{figure*}[t]
\centering
\includegraphics[width=0.75\textwidth]{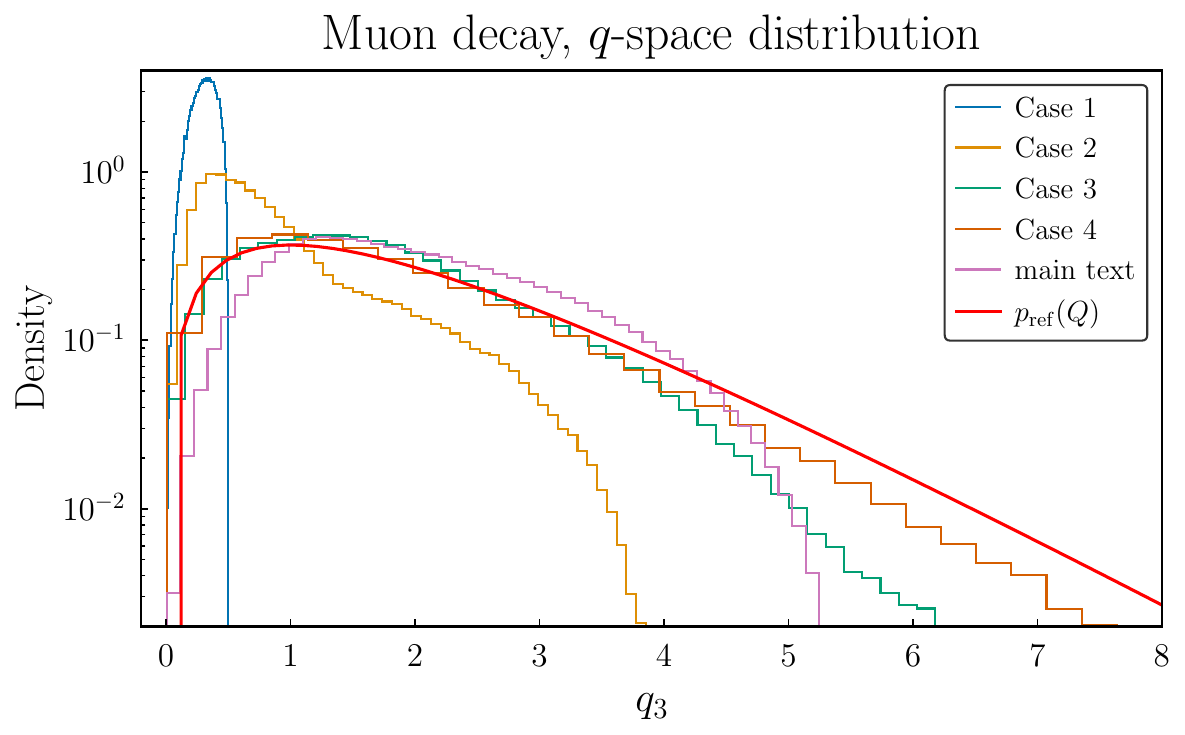}
\caption{Training data in $q$-space for the different data augmentation strategies in Fig.~\ref{fig:vary_Nmult}.}
\label{fig:vary_Nmult_qspace}
\end{figure*}

Fig.~\ref{fig:vary_Nmult} shows the effects of different data augmentation strategies on the kinematic distributions of particle 3 from the muon decay example of Sec.~\ref{sec:Muon}. We compare the default $N_{\rm mult} = 1$ case described in the main text to four alternative cases:
\begin{enumerate}
    \item no augmentation or transformation (i.e. interpreting $p$-space vectors directly as $q$-space vectors);
    \item a 2-fold augmentation $N_{\rm mult} = 2$;
    \item a larger augmentation $N_{\rm mult} = 10$;
    \item a continuous transformation where each $p$-space point is boosted and rescaled by a \emph{different} conformal transformation $(\mathbf{b},x)$.
\end{enumerate}
All models were trained with the same total $q$-space training set size of 500,000, with all other hyperparameters the same as described above. The energy distributions are clearly a poor match to the target, with all but the default strategy showing a pronounced spurious excess near the kinematic endpoint of $E_3 = 0.5$. By contrast, more augmentation in $q$-space (cases 3 and 4 above) gives better angular distributions, because the anisotropies induced from a single choice of $\mathbf{b}$ are washed out in $q$-space. We can see this in Fig.~\ref{fig:vary_Nmult_qspace}, where the $q$-space training data for cases 3 and 4 are already very close to the forward process endpoint $p_{\rm ref}(Q)$.

\begin{figure*}[t]
\centering
\includegraphics[width=0.7\textwidth]{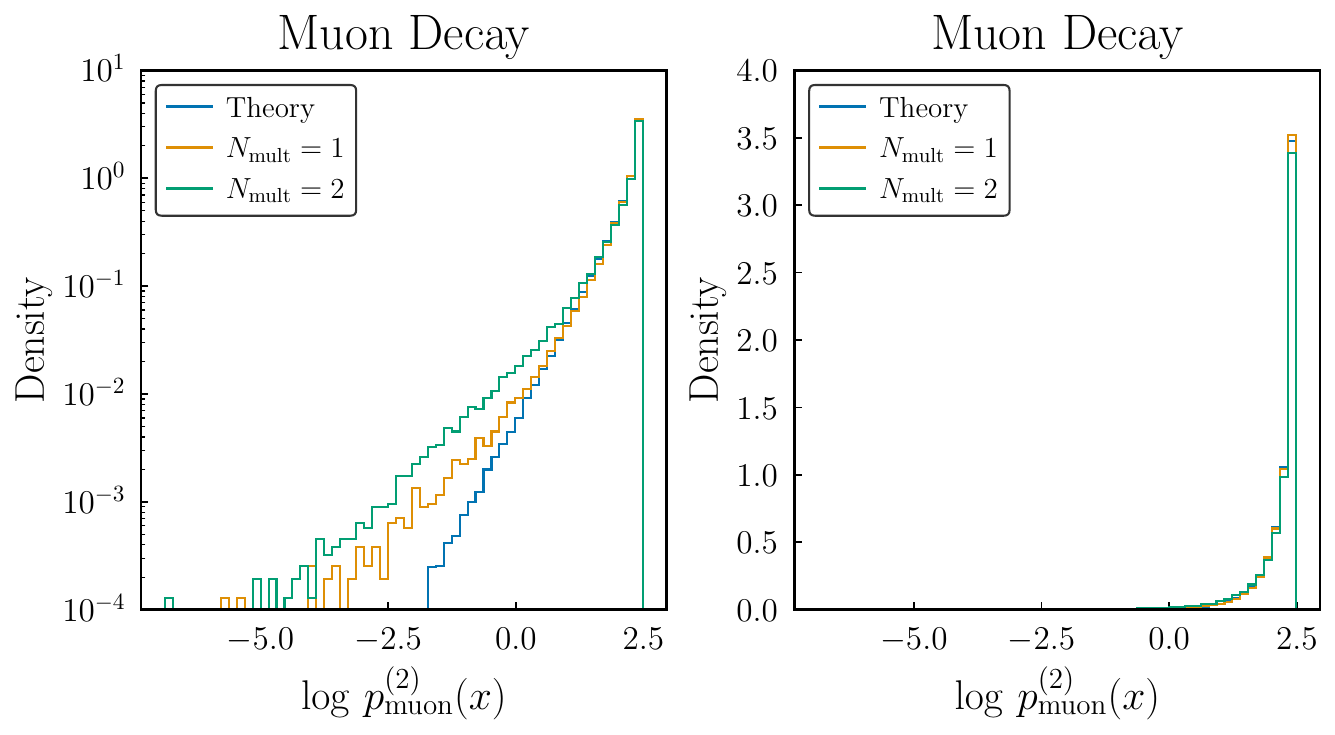}
\caption{
Distributions of the logarithm of the theoretical Dalitz plot PDF for the true muon decay distribution and for generated distributions with $N_{\rm mult} = $ 1 or 2. The model with $N_{\rm mult} = 2$ is a worse match to the true distribution even for the high-probability events.}
\label{fig:Dalitz_LogPDF_KS_Plots_Full}
\end{figure*}

\begin{figure*}[h]
\centering
\includegraphics[width=0.99\textwidth]{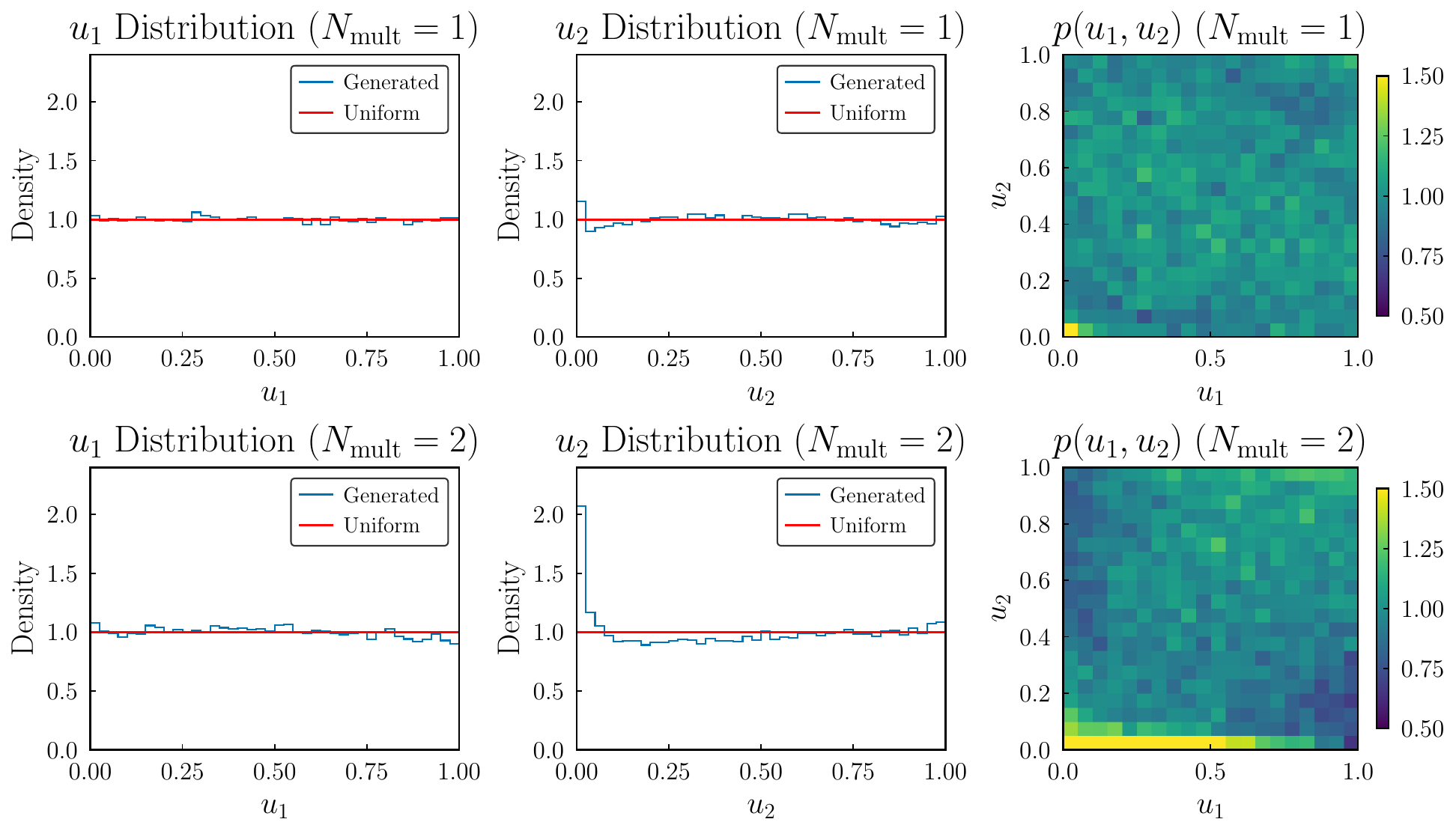}
\caption{
Rosenblatt transformations as in Fig.~\ref{fig:Rosenblatt_Plots}, but for diffusion models trained with $N_{\rm mult} = $ 1 or 2. The non-uniformity of both $u_1$ and $u_2$ is evident for $N_{\rm mult} = 2$.
}
\label{fig:Rosenblatt_Plots_Full}
\end{figure*}

To quantify the effect of data augmentation on the energy distributions, Figs.~\ref{fig:Dalitz_LogPDF_KS_Plots_Full} and \ref{fig:Rosenblatt_Plots_Full} show the same 1D and 2D distributions as in the main text, comparing case 2 above with the default strategy. Table~\ref{tab:Wasserstein_compare} then presents the Wasserstein-1 distances between various 1D projections of the results from the two cases above and the theoretical distribution. Notably, although case 2 leads to a substantial improvement in the output's isotropy (see especially the improvement in the distribution of $\cos\theta_3$), this comes at the price of worse agreement on the energy distributions (as evidenced by the $E_I$ and $u_j$ distributions).

\begin{table*}[t]
    \centering
%    \scriptsize
    \setlength{\tabcolsep}{4pt}
    \begin{tabular*}{0.5\textwidth}{@{\extracolsep{\fill}}lcc}
        Observable & $N_{\rm mult}=1$ & $N_{\rm mult}=2$ \\
        \hline
        $E_1$ & $\mathbf{0.73\times10^{-3}}$ & $2.4\times10^{-3}$ \\
        $E_2$ & $\mathbf{0.35\times10^{-3}}$ & $1.2\times10^{-3}$ \\
        $E_3$ & $\mathbf{1.6\times10^{-3}}$ & $5.7\times10^{-3}$ \\
        $\ln\left(p_{\rm muon}^{(2)}\right)$ & $\mathbf{17\times10^{-3}}$ & $78\times10^{-3}$ \\
        $\cos\theta_3$ & $60\times10^{-3}$ & $\mathbf{3.8\times10^{-3}}$ \\
        $\cos\theta_{\rm EP}$ & $\mathbf{2.2\times10^{-3}}$ & $3.8\times10^{-3}$ \\
        $\phi_{\rm EP}$ & $19\times10^{-3}$ & $\mathbf{7.4\times10^{-3}}$ \\
        $u_1$ & $\mathbf{1.9\times10^{-3}}$ & $7.4\times10^{-3}$ \\
        $u_2$ & $\mathbf{3.0\times10^{-3}}$ & $10\times10^{-3}$ \\
        \hline
    \end{tabular*}
    \caption{The Wasserstein-1 distances between various projections of the event distributions generated using $N_{\rm mult}=1$ or $2$ and using the theoretical event distribution; all variables are defined in the main text. Note that Wasserstein distances carry units and, correspondingly, scale with the variation of the underlying variable, so comparisons between distinct rows must be made with caution; the two columns, however, can be safely compared. The lower value in each row is bolded.}
    \label{tab:Wasserstein_compare}
\end{table*}

An alternative way to summarize the trade-offs of different data augmentation strategies is through aggregate metrics such as Fr\'echet distances in suitably chosen feature spaces and classifier two-sample tests. In the present setting, however, the choice of features is itself nontrivial. Lorentz-invariant features such as pairwise invariant masses correctly quotient out overall rotations, but are insensitive to the angular anisotropy visible in Fig.~\ref{fig:muon_angles}. Conversely, learned embeddings built from the full momenta can respond to those angular artifacts as well as to other representation-dependent details. We therefore use these metrics only as a compact cross-check. Specifically, we train a small autoencoder on truth events only and use its latent space as a shared feature embedding for all five models in Fig.~\ref{fig:vary_Nmult}. 

Table~\ref{tab:muon_metric_compare} reports Fr\'echet distances and classifier two-sample test (C2ST) areas-under-the-curve (AUCs) in both invariant and learned feature spaces, together with Kolmogorov-Smirnov (KS) statistics for the $E_3$ and $\cos\theta_3$ marginals. For the learned-feature quantities, we repeat the shared-autoencoder training over two seeds, and for the classifier results we additionally repeat the train/test split and quote the mean and standard deviation of a linear C2ST AUC. The pattern mirrors the trade-off visible in the histograms of the 1D marginal distributions: the default model gives the best invariant-space and energy-space agreement, while larger or continuous augmentation improves the angular distribution and tends to score better in the learned embedding. This dependence on representation is precisely why, for these low-dimensional examples, we view the aggregate metrics as secondary to the comparisons of the marginal distributions via the Wasserstein distance.

\begin{table*}[t]
    \centering
    \scriptsize
    \setlength{\tabcolsep}{4pt}
    \begin{tabular*}{\textwidth}{@{\extracolsep{\fill}}lcccccc}
        & FID$_{\rm inv}$ & FID$_{\rm AE}$ & AUC$_{\rm inv}$ & AUC$_{\rm AE}$ & KS$(E_3)$ & KS$(\cos\theta_3)$ \\
        \hline
        default & $\mathbf{4.8 \times 10^{-5}}$ & $0.052 \pm 0.002$ & $\mathbf{0.500 \pm 0.008}$ & $0.527 \pm 0.011$ & $\mathbf{0.0270}$ & $0.0667$ \\
        case 1: none & $1.8 \times 10^{-4}$ & $0.076 \pm 0.003$ & $0.503 \pm 0.004$ & $0.545 \pm 0.007$ & $0.0353$ & $0.0551$ \\
        case 2: $N_{\rm mult}=2$ & $6.9 \times 10^{-5}$ & $0.029 \pm 0.001$ & $0.486 \pm 0.005$ & $0.511 \pm 0.007$ & $0.0355$ & $0.0214$ \\
        case 3: $N_{\rm mult}=10$ & $9.2 \times 10^{-5}$ & $0.033 \pm 0.001$ & $0.501 \pm 0.008$ & $0.509 \pm 0.012$ & $0.0480$ & $0.0175$ \\
        case 4: cont. & $1.8 \times 10^{-4}$ & $\mathbf{0.019 \pm 0.002}$ & $0.513 \pm 0.009$ & $\mathbf{0.494 \pm 0.010}$ & $0.0480$ & $\mathbf{0.0143}$ \\
        \hline
    \end{tabular*}
    \caption{Compact comparison of the five muon-decay models shown in Fig.~\ref{fig:vary_Nmult}. All quantities are evaluated on the same 2000-event held-out sample. ``FID$_{\rm inv}$'' denotes a Fr\'echet distance computed on Lorentz-invariant features, while ``FID$_{\rm AE}$'' uses the latent space of a shared autoencoder trained only on truth events; for the latter we quote the mean and standard deviation over two independent shared-AE trainings. ``AUC$_{\rm inv}$'' and ``AUC$_{\rm AE}$'' are linear classifier two-sample test AUCs in the same two feature spaces, with mean and standard deviation taken over repeated random train/test splits and classifier seeds; values closer to $0.5$ indicate better agreement with the target distribution. The final two columns give KS statistics for the energy and angular marginals most visibly affected by the augmentation choice. Lower is better for all metrics shown, except that the AUC values are optimal at $0.5$; the best entry in each column is bolded. For case 1, one non-finite sample out of 2000 generated events is removed before computing these aggregate metrics.}
    \label{tab:muon_metric_compare}
\end{table*}

\subsubsection{Diffusion schedule}

Here we investigate the effect of changing the diffusion schedule. We take the $q\bar{q}g$ distribution from Sec.~\ref{sec:qqg} as an example. The diffusion schedule used in the main text has $T = 500$ steps. We consider extending this to $T = 2000$ steps with a Gaussian diffusion phase of $t_{\rm gaus} = 50$ steps at the beginning of the forward process. We also consider the no-augmentation strategy of interpreting the $p$-space data directly in $q$-space.

\begin{figure*}[t]
\centering
\includegraphics[width=0.9\textwidth]{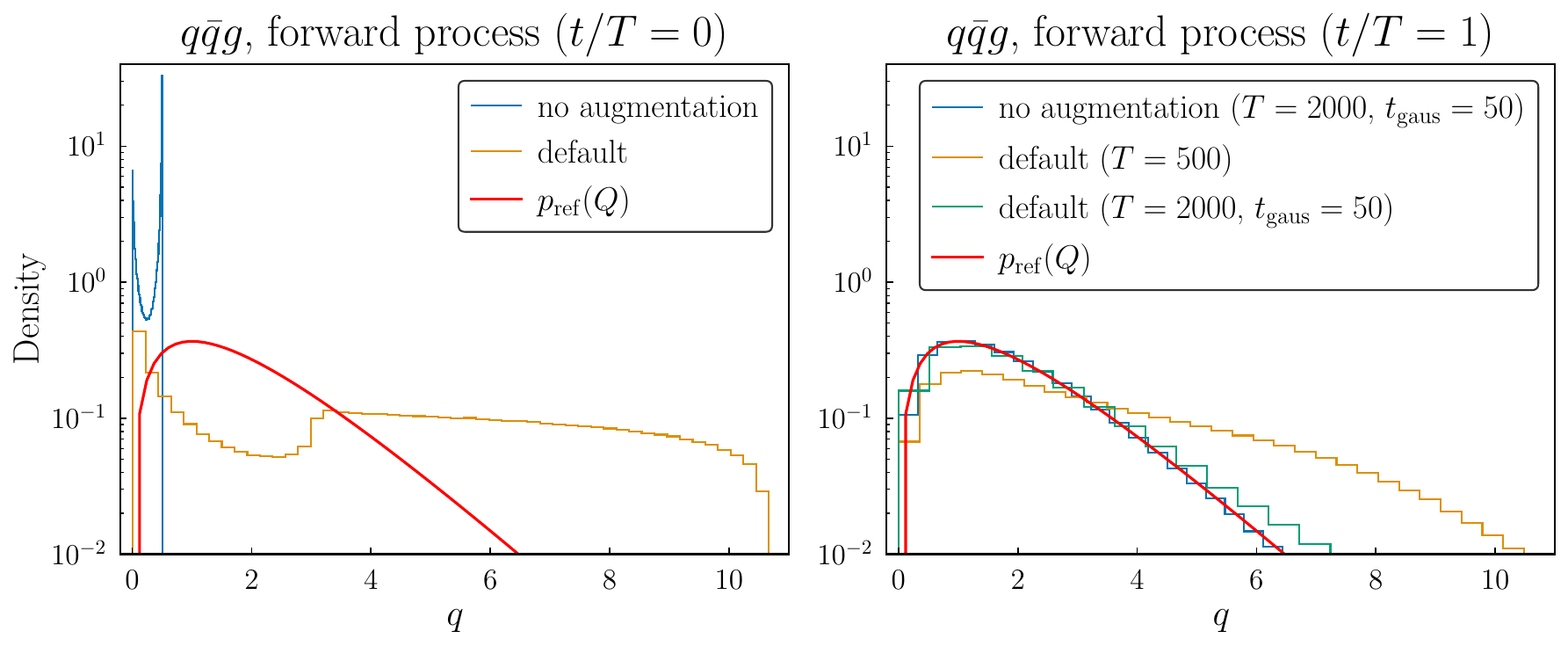}
\caption{The $q$-space distributions at the start (\emph{left}) and end (\emph{right}) of the forward process, for different diffusion schedules and data augmentation strategies.}
\label{fig:vary_sched_forward}
\end{figure*}

\begin{figure*}[t]
\centering
\includegraphics[width=0.99\textwidth]{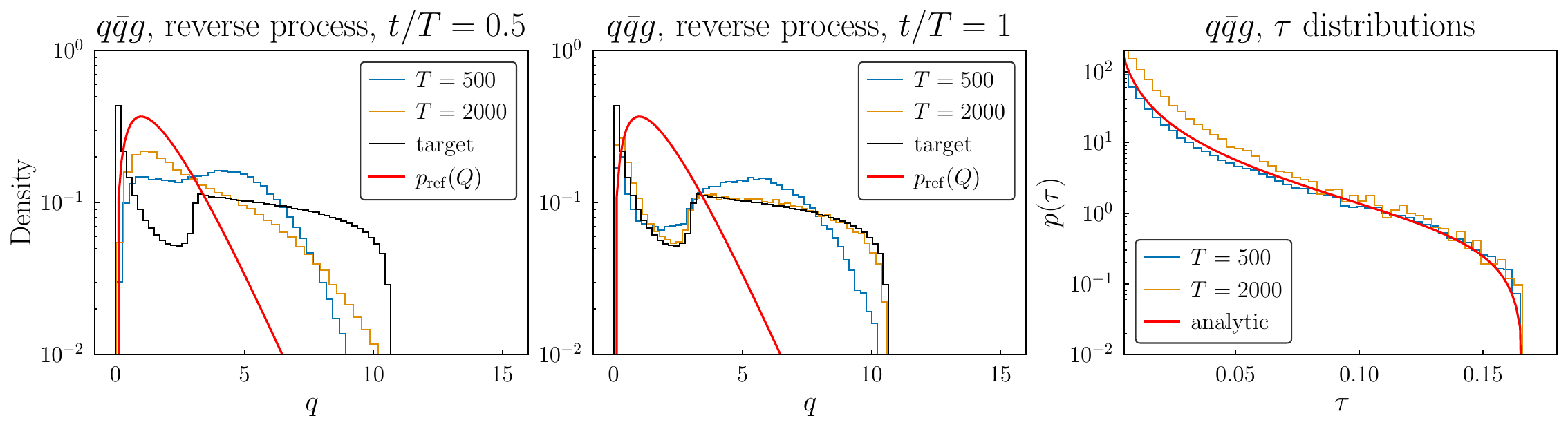}
\caption{Reverse trajectory for the $N_{\rm mult} = 1$ data augmentation strategy. The left panel shows the $q$-space distributions halfway through the reverse process $(t/T = 0.5)$, while the center panel shows the endpoint of the reverse process. The right panel shows the $\tau$ distribution analogous to Fig.~\ref{fig:taudist}.}
\label{fig:vary_sched_rev}
\end{figure*}

Fig.~\ref{fig:vary_sched_forward} shows the beginning and end of the forward process for these three different schedules and datasets. The shorter schedule does not fully equilibrate in $q$-space, while the longer schedule (for both training set choices) is a much better fit to the target distribution $p_{\rm ref}(Q)$. Fig.~\ref{fig:vary_sched_rev} shows the reverse process trajectory at normalized time $t/T = 0.5$ and at the endpoint $t/T = 1$ for the default $N_{\rm mult} = 1$ data augmentation. In $q$-space, it is clear that the longer diffusion time leads to a better match to the target distribution in $q$-space (center panel). However, this does not translate directly to better performance in physical phase space; the right panel shows that the $\tau$ distribution is a \emph{worse} match to the analytic distribution for the longer diffusion time. This example reinforces the point that $q$-space is explicitly unphysical; many correlations in $q$-space are spurious and are projected out when mapping to phase space, and thus a model which is encouraged to match $q$-space distributions exactly may actually not be desirable. This may also explain the observations from App.~\ref{app:augmentation} above that more data augmentation leads to poorer energy distributions when sampled from the reverse process. Larger $N_{\rm mult}$ makes the initial $q$-space distribution closer to the equilibrium distribution, but at the cost of washing out important correlations in $p$-space. 

\begin{figure*}[t]
\centering
\includegraphics[width=0.9\textwidth]{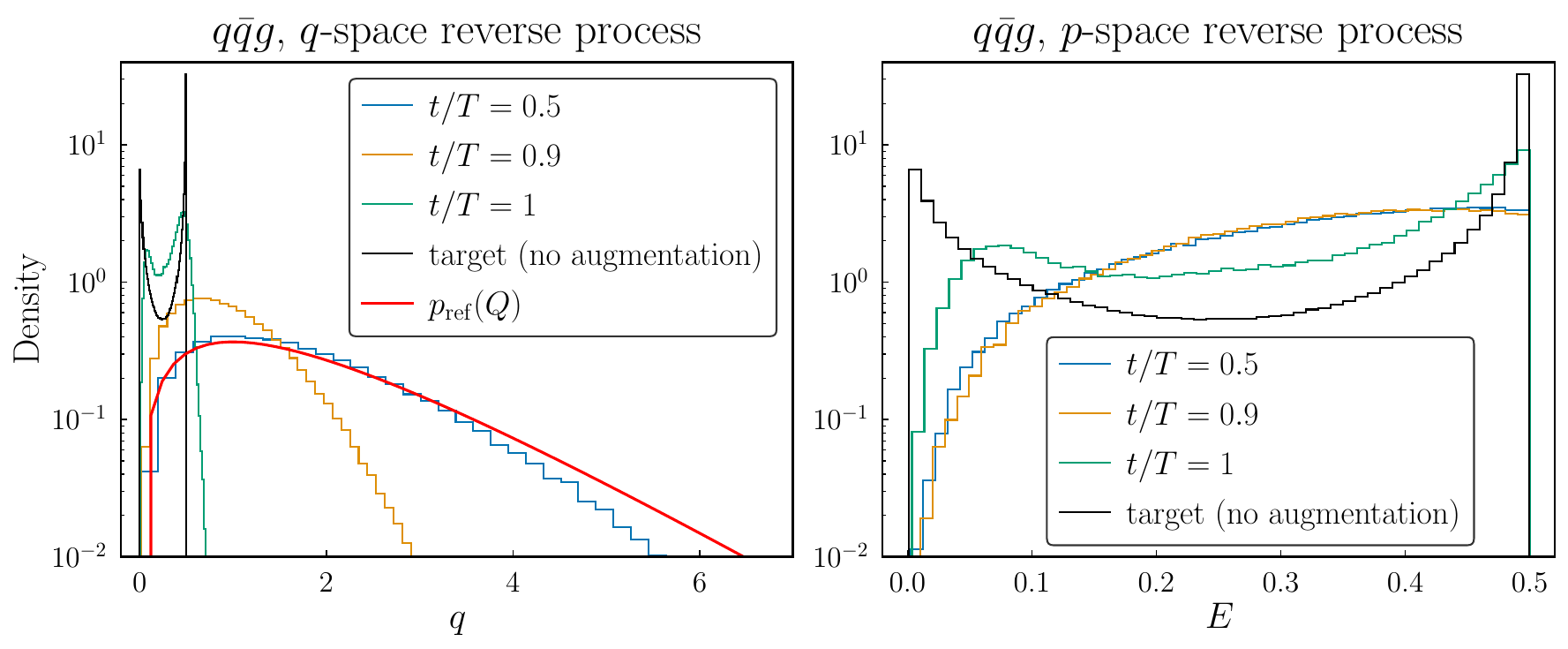}
\caption{Snapshots from the reverse process in $q$-space (\emph{left}) and $p$-space (\emph{right}), for the case of $q\bar{q}g$ data with no augmentation in $q$-space.}
\label{fig:rev_nofluff}
\end{figure*}

We see this effect even more clearly in Fig.~\ref{fig:rev_nofluff}, which shows snapshots from the reverse trajectory in both $q$-space and $p$-space for the no-augmentation strategy. Motion in $q$-space towards the target distribution is roughly continuous, but the $p$-space distributions do not deviate significantly from the uniform distribution until the last 10\% of the reverse trajectory. This is because most of the reverse process is spent ``bringing in the tails'' of the $q$-space distribution, which does not actually map to the desired correlations in $p$-space.

\subsection{Higher-dimensional distributions}
\label{app:high-dim-qspace}
\paragraph{Network.}
 For distributions with $N \geq 3$ particles, we replace the MLP score network with a \textsc{Point Edge Transformer} (PET) as described in \cite{OmniLearned} due to the architecture's track record for jet tasks and its ability to generalize beyond what it was originally designed for \cite{Mikuni:2025ocp,Elsharkawy:2026kwp}. The PET is a transformer that treats each $q$-space event as a point cloud of $N$ particles: it is inherently permutation-equivariant and can handle arbitrary-length inputs.

Our PET score network uses the ``small'' configuration described in \cite{OmniLearned} and given by 8 transformer blocks, 2 local attention layers, 8 attention heads, a latent dimension of 128, and $K=9$ nearest neighbors for the local attention block. The generator head maps the transformer output back to a per-particle 3-dimensional noise prediction, conditioned on diffusion time via a sinusoidal embedding.

\paragraph{$q$-space diffusion in high dimensions.}
In high dimensions, computing the divergence of the network for each jet constituent becomes prohibitively expensive. One way around this is the well-known Hutchinson trace estimator which approximates $\nabla \cdot \mathbf{s}_\theta$ as $\mathbb{E}_\mathbf{v}[\mathbf{v}^\top (\nabla_Q \mathbf{s}_\theta), \mathbf{v}]$ for random probe vectors $\mathbf{v}$, reducing the cost from $3N$ backward passes to one. However, this estimator still requires second-order derivatives through the network (requiring $\sim2-5\times$ more compute). We also find empirically that it is unstable to train. 

\begin{figure*}[t]
\centering
\includegraphics[width=0.99\textwidth]{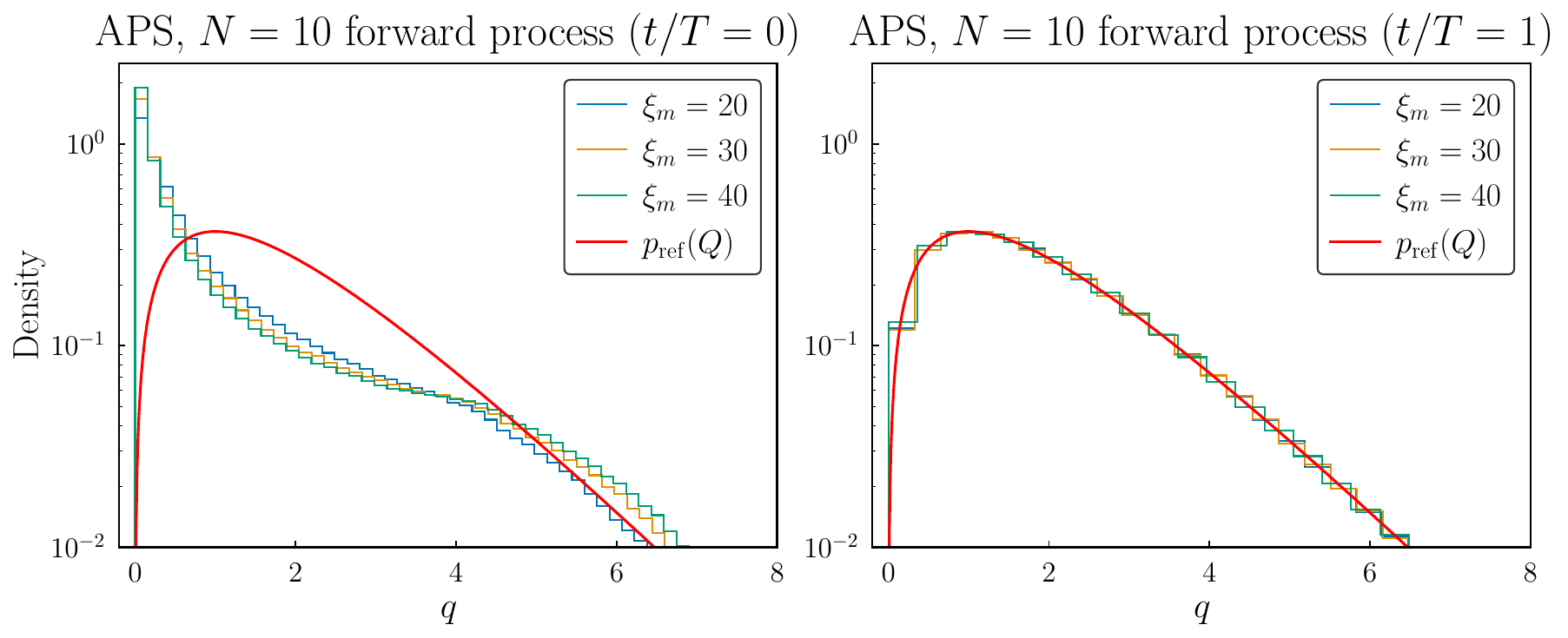}
\caption{Forward process start and end distributions in $q$-space for the $N = 10$ APS distributions.}
\label{fig:app_APS_forward}
\end{figure*}

Instead we use a version of denoising score matching~\cite{Ho:2020ddpm}. To train the score network $\mathbf{s}_\theta$, we first pre-compute the forward process up to a randomly sampled time $t$, obtaining $Q_t$. We then take one more forward step to produce
\begin{equation}
Q_{t+1} = Q_t + \gamma_t \nabla \log p_{\rm ref}(Q_t) + \sqrt{2\gamma_t}Z_t,
\end{equation}
where $Z_t \sim \mathcal{N}(0,I)$. This is to say $p(Q_{t+1} \mid Q_t) \sim \mathcal{N}\big(\mu, 2\gamma_t\cdot I\big)$ with $\mu=Q_t + \gamma_t \nabla \log p_{\rm ref}(Q_t)$. Given the Gaussianity of the conditional distribution, we can show
\begin{equation}
 \nabla_{Q_{t+1}}\log p(Q_{t+1} \mid Q_t) = -\frac{Q_{t+1} - Q_t - \gamma_t \nabla \log p_{\rm ref}(Q_t)}{2\gamma_t} \equiv -\frac{Z_t}{\sqrt{2\gamma_t}}.
\end{equation}
where we use the forward equation in the second equality. Importantly, \cite{Vincent2011} shows that,
\begin{equation}
        \mathbb{E}_{Q_t \sim p_t}\Big[||\mathbf{s}_\theta(Q_{t+1}, t) - \nabla_{Q_{t+1}} \log p_t(Q_{t+1})||^2\Big] = \mathbb{E}_{Q_t, Q_{t+1}}\Big[||\mathbf{s}_\theta(Q_{t+1}, t) - \nabla_{Q_{t+1}} \log p(Q_{t+1} \mid Q_t)||^2\Big] + C
\end{equation}
for any $p_t(Q_{t+1})$. That is, if the network learns the gradient of the transition kernel, it is also learning the score $\nabla_{Q_{t+1}}\log p_t(Q_{t+1}) \equiv\mathbf{s}_\theta $. However, the $1/\sqrt{2\gamma_t}$ pre-factor causes the loss to vary by orders of magnitude across diffusion time steps. So we instead reparameterize the network to predict the noise itself: $\boldsymbol{\varepsilon}_\theta(Q_{t+1}, t) \approx Z_t$, with the loss
\begin{equation}
\mathcal{L}_{\rm DSM\text{-}\varepsilon} = \mathbb{E}_{t,, Z_t}\Big[\big|\big|\boldsymbol{\varepsilon}_\theta(Q_{t+1}, t) - Z_t\big|\big|^2\Big].
\end{equation}
Since $Z_t \sim \mathcal{N}(0,I)$ is unit-variance regardless of $\gamma_t$, this loss remains $\mathcal{O}(1)$ throughout training. At generation time, the predicted noise is converted back to a score via $\mathbf{s}_\theta = -\boldsymbol{\varepsilon}_\theta / \sqrt{2\gamma_t}$ and substituted into the reverse process. The network implicitly learns both the deterministic drift $\gamma_t \nabla \log p_{\rm ref}(Q_t)$ and the stochastic contribution $\sqrt{2\gamma_t}Z_t$ that together compose each forward step.
\paragraph{$p$-space diffusion.}
As a comparison to the $q$-space approach above, we also train a standard DDPM model \cite{Ho:2020ddpm} that operates directly on the 3-momentum vectors $P =\{\p_I\}$.  That is, the forward process is given by,
\begin{equation}
P_{t+1} =  \sqrt{1-2\gamma_t} P_t + \sqrt{2\gamma_t}Z_t, \qquad Z_t\sim\mathcal{N}(0,I),
\end{equation}
The network is again reparameterized to predict noise, $\boldsymbol{\varepsilon}_\theta(P_t, t)\approx Z_t$, with the same loss as above. Generation also follows the standard time-reversed Langevin dynamics, with the trained model acting as the time-dependent score. 

\paragraph{Flow matching models.} For flow matching, we use the same \textsc{Point-Edge-Transformer} architecture described above, with the only change being the loss function. The loss for both $p$-space and $q$-space is given in Eq.~(\ref{eq:qflow_loss}).

\paragraph{Training configuration.}
Similar to $N = 3$ case, the diffusion schedule consists of $T=5000$ steps with a linear gamma schedule from $\gamma_{\min}=0.001$ to $\gamma_{\max}=0.005$. The first 100 steps are the Gaussian phase described in Sec.~\ref{sec:Antenna} with $\gamma = 10^{-4}$. We use a single $q$-space augmentation ($N_{\mathrm{mult}} = 1$) per training sample, drawing from 900,000 training events (withholding 100,000 events for test and/or validation). Fig.~\ref{fig:app_APS_forward} shows the distributions of $q_I$ at the start and end of the forward process, demonstrating good convergence to the forward process endpoint $p_{\rm ref}(Q)$.

Training for all models (both diffusion and flow matching) uses the \texttt{AdamW} optimizer with $\beta_1=0.95$, $\beta_2=0.99$, weight decay of $0.01$, and gradient clipping at norm 1.0. The learning rate is $10^{-3}$ with a \textsc{OneCycle} scheduler \cite{OneCycleLR}. We train with batch size of 1024 on 4 NVIDIA L40S GPUs for 6 hours and take the state of the network after that time as the trained model.

\section{Derivation of $\tau$ distributions on uniform phase space}\label{app:tauder}

In this appendix, we present the derivation of the distribution of the minimum four-vector dot product $\tau = \min\{p_I\cdot p_J\}$ on uniform phase space, for both $N=3$ and in the large-$N$ limit.  Starting with $N=3$, we note that phase space $\Pi_3$ can be expressed as (dropping normalization factors)
\begin{align}
d\Pi_3 \propto ds_{12}\,ds_{23}\, ds_{13}\,\delta(1-s_{12}-s_{23}-s_{23})=ds_{12}\,ds_{23}\, \Theta(1-s_{12}-s_{23})\,,
\end{align}
assuming a uniform distributions of orientations on the celestial sphere $S^2$.  The distribution of the minimum dot product can be calculated from
\begin{align}
p^{(3)}_{\rm uniform}(\tau) &\propto \int ds_{12}\, ds_{23}\,\Theta(1-s_{12}-s_{23})\,\Theta(s_{23}-s_{12})\Theta(1-s_{12}-s_{23}-s_{12})\delta\left(\tau - \frac{s_{12}}{2}\right)\nonumber\\
&\propto \frac{1}{6}-\tau\,,
\end{align}
where the $\Theta$ functions select $s_{12}$ as the smallest invariant mass.

In the limit of a large number of identical particles, the single-particle four-momentum distribution thermalizes, and is simply a Boltzmann distribution:
\begin{align}
p^{(N)}\left(p^\mu_I\right) = \frac{2N^2}{\pi}\,e^{-2NE_I}\,,
\end{align}
where $E_I$ is the particle's energy in the event's CM frame. This distribution is normalized on Lorentz invariant, on-shell single particle phase space,
\begin{align}
1 = \int d^4p_I\, \delta(p_I^2)\, p^{(N)}\left(p_I\right) = \int \frac{d^3{\bf p}_I}{2E_I}\,\frac{2N^2}{\pi}\,e^{-2NE_I}\,.
\end{align}
Its mean is $\langle E\rangle = 1/N$, where the total CM energy is 1.

Given this Boltzmann distribution, the distribution of the four-vector dot product of momenta between two particles 1 and 2, $s\equiv p_1\cdot p_2$, is
\begin{align}
p^{(N)}(s) &= \int d^4p_1\,\delta(p_1^2)\,d^4p_2\, \delta(p_2^2)\, p^{(N)}\left(p^\mu_1\right)\, p^{(N)}\left(p^\mu_2\right)\,\delta\left[s-E_1E_2(1-\cos\theta_{12})\right]\\
&=8N^4\int dE_1\, dE_2\, d\cos\theta_{12}\, E_1\,E_2\, e^{-2N(E_1+E_2)}\,\delta\left[s-E_1E_2(1-\cos\theta_{12})\right]\nonumber\\
&=4N^3\sqrt{2s}\,K_1\left(
2N\sqrt{2s}
\right)\,,
\nonumber
\end{align}
where $K_1(x)$ is a modified Bessel function.  Its cumulative distribution is
\begin{align}
\Sigma^{(N)}(s) = 1-(2N)^2 s\,K_2\left(
2N\sqrt{2s}
\right)\,,
\end{align}
where $K_2(x)$ is another modified Bessel function.

Now, in the large-$N$ thermalized limit where the four energy-momentum constraints can be ignored, dot products between any pair of particles are independent and identically distributed. Further, there are ${N\choose 2}\sim N^2/2$ such pairs, in the large-$N$ limit.  It follows that the cumulative distribution of the minimum dot product, the observable $\tau$ follows simply as
\begin{align}
\Sigma^{(N)}_{\rm min}(\tau) = 1-\left(
1-\Sigma(s=\tau)
\right)^{N^2/2} \to 1-\exp\left[
-\frac{N^2}{2}\left(
1-(2N)^2 \tau\,K_2(
2N\sqrt{2\tau})
\right)
\right]\,.
\end{align}
The distribution of the minimum dot product follows by taking a derivative:
\begin{align}
p^{(N)}_\text{uniform}(\tau) &= \frac{d}{d\tau}\Sigma^{(N)}_{\rm min}(\tau) \\
&= 2N^5\sqrt{2\tau}\,K_1\left(
2N\sqrt{2\tau}
\right)\,\exp\left[
-\frac{N^2}{2}\left(
1-(2N)^2 \tau\,K_2(
2N\sqrt{2\tau})
\right)
\right]\,.\nonumber
\end{align}
We expect this distribution to capture the correct behavior at \emph{small} $\tau$, away from the kinematic endpoint where energy-momentum conservation imposes $\mathcal{O}(1/N)$ correlations among the $N$ particles. This agreement with \texttt{RAMBO}-sampled events is demonstrated in Fig.~\ref{fig:SARGE_taudist}.

\bibliography{jetdiffusion.bib}

\end{document}